%% file: main.tex
\documentclass[11pt]{article}
\usepackage{graphicx,amsmath,amssymb}
\usepackage[margin=1.25in]{geometry}
\usepackage[usenames,dvipsnames]{color}
\usepackage[T1]{fontenc}
\usepackage{url}
\usepackage{cite}
\usepackage{xcolor}
\usepackage{lineno}
\usepackage{subfigure}
\usepackage{xspace}
\usepackage[colorlinks = true,
            linkcolor = blue,
            urlcolor  = blue,
            citecolor = blue,
            anchorcolor = blue]{hyperref}

\usepackage[section]{placeins}

\newcommand\pubnumber{Transcendental Preprint}
\newcommand\pubdate{\today}

\def\Title#1{\begin{center} {\LARGE #1 } \end{center}}
\def\Author#1{\begin{center}{ \sc #1} \end{center}}

\newcommand\pubblock{\rightline{\begin{tabular}{l} \pubnumber\\
         \pubdate \end{tabular}}}
\newenvironment{Abstract}{\begin{quotation} \begin{center}
                       ABSTRACT
     \end{center}\bigskip  }{\end{quotation}}

\input workshopsymbols.tex

\newcommand\snowmass{\begin{center}\rule[-0.2in]{\hsize}{0.01in}\\\rule{\hsize}{0.01in}\\
\vskip 0.1in Submitted to the  Proceedings of the US Community Study\\ 
on the Future of Particle Physics (Snowmass 2021)\\ 
\rule{\hsize}{0.01in}\\\rule[+0.2in]{\hsize}{0.01in} \end{center}}

\begin{document}

\pubblock

\Title{Searches for Long-Lived Particles at the Future FCC-ee}

\bigskip 

\input{authorlist.tex}

\medskip

\begin{Abstract}

\noindent The electron-positron stage of the Future Circular Collider, FCC-ee, is a frontier factory for Higgs, top, electroweak, and flavour physics. 
It is designed to operate in a 100\,km circular tunnel built at CERN, and will serve as the first step towards $\geq$ 100\,TeV proton-proton collisions. In addition to an essential and unique Higgs program, it offers powerful opportunities to discover direct or indirect evidence of physics beyond the Standard Model.

Direct searches for long-lived particles at FCC-ee could be particularly fertile in the high-luminosity $Z$ run, where $5\times 10^{12}$ $Z$ bosons are anticipated to be produced for the configuration with two interaction points. The high statistics of Higgs bosons, $W$ bosons and top quarks in very clean experimental conditions could offer additional opportunities at other collision energies. Three physics cases producing long-lived signatures at FCC-ee are highlighted and studied in this paper: heavy neutral leptons (HNLs), axion-like particles (ALPs), and exotic decays of the Higgs boson. These searches motivate out-of-the-box optimization of  experimental conditions and analysis techniques, which could lead to improvements in other physics searches. 

\end{Abstract}

\snowmass
\def\thefootnote{\fnsymbol{footnote}}
\setcounter{footnote}{0}

\clearpage
\tableofcontents
\clearpage

\section{Introduction}\label{intro}
\emph{Editor: Rebeca Gonzalez Suarez}

The Standard Model (SM) of particle physics is a mature and consistent theory that, after the observation of the Higgs boson, still fails to explain important experimental observations such as dark matter (DM), neutrino masses, or the baryon asymmetry of the universe (BAU), among others. Theoretical aspects of the SM, including the origins of the electroweak scale, the spectrum of fermions masses, or flavor patterns also await explanation. These questions may be answered by the direct observation of new particles and phenomena, or by measuring deviations from SM predictions. This is a chief motivation for new colliders that can push both the energy and intensity frontiers.  

Long-lived particles (LLPs) are new, beyond the SM (BSM) states that travel a substantial distance between creation and decay in collider systems, presenting  distinct experimental signatures~\cite{Alimena:2019zri}. LLPs feature in many BSM models and could provide answers to central questions in particle physics and beyond. The lifetime of a particle depends mostly on its mass and couplings, and so feebly-interacting particles (FIPs), with couplings to the SM particles several orders of magnitude smaller than the SM couplings, are often glaring examples of LLPs. 

The experimental signatures of LLPs are particularly interesting. In contrast to promptly decaying particles, LLPs can decay after flying some distance from the primary interaction point. This produces a displaced vertex, with decay products including charged and neutral SM particles (e.g. charged leptons, light neutrinos, and pions). This kind of displaced signature is most commonly associated with LLPs. Other models predict disappearing LLPs giving rise to ``short'' or ``broken'' tracks; some are ``stopped'' or delayed; or produce unusual jets, such as ``dark showers''. Such a variety of experimental signatures is very different from the usual SM processes studied at colliders, and any of these signatures would, if observed, constitute a striking ``smoking gun'' of new physics. In hadron collider environments, standard trigger and reconstruction techniques are often unable to recognize LLP signatures and their study requires dedicated techniques and experiments.

The Future Circular Collider (FCC) program is a design study for a post-LHC particle accelerator at CERN following the priorities set by the 2020 Update of the European Strategy for Particle Physics~\cite{CERN-ESU-015}. The first stage of the FCC design study (FCC-ee) is a high-luminosity, high-precision lepton collider with the goal of better understanding the electroweak (EW) sector, especially the Higgs boson. In addition to a robust program in its own right, FCC-ee will also act as a possible precursor to a high-energy hadron collider (FCC-hh), located in the same tunnel and complementary to it~\cite{FCC-CDR2}.

Though FCC-ee will be a high precision exploration tool, it also opens the possibility of directly discovering new physics~\cite{Chrzaszcz:2021nuk}. In particular, a future FCC-ee program has an exciting potential for exploring LLPs. The large integrated luminosity of the FCC-ee run around the $Z$ pole, producing $5\times10^{12}$ $Z$ bosons (Tera-$Z$ run), will facilitate direct searches for LLPs that could be closely linked to neutrino masses, explain the BAU, be sound DM candidates, or all at the same time. In the following, three central physics cases are discussed: heavy neutral leptons (HNLs)~\cite{Abdullahi:2022jlv} in the context of the Phenomenological Type I Seesaw model, axion-like particles (ALPs), and exotic Higgs boson decays. In Section~\ref{sec:theory}, the theoretical landscapes of these three physics cases are outlined. Section~\ref{sec:exp} discusses the experimental outlook. In particular, the common simulation details (Section~\ref{sec:analysis_setup}), the experimental aspects of HNLs (Section~\ref{sec:analysis_hnl}), ALPs (Section~\ref{sec:analysis_alp}), and exotic Higgs boson decays (Section~\ref{sec:analysis_higgs}) at FCC-ee, and considerations for additional detectors for LLPs at FCC-ee (Section~\ref{sec:analysis_additionalDetectors}) are covered. Finally, the summary and conclusions are presented in Section~\ref{sec:summary}.

\input{theory}

\input{experiment}

\section{Summary and Conclusions} \label{sec:summary}
In this paper, we discuss three key BSM cases at the future FCC-ee that experimentally can display long-lived signatures: heavy neutral leptons, axion-like particles, and exotic Higgs boson decays. 
While FCC-ee is primarily envisioned as a precision collider, the discussed scenarios are examples of direct searches that could be performed and could answer central questions of particle physics. 

The three cases are carefully discussed from a theoretical perspective, representing the state-of-the-art and current best expected limits. Simulation studies are then presented for HNLs and ALPs. These two BSM cases can present displaced signatures: a displaced vertex for the former, and a displaced photon pair in the latter. 

Different HNL signals---as well as a limited collection of background processes---are generated, kinematic variables are explored, and a first possible set of requirements is presented to isolate the signal from the SM backgrounds. Possible kinematic variables that could characterize an HNL as Dirac or Majorana are also explored experimentally.

For ALPs, signals are generated and validated and some key distributions are presented. 

The work presented here can be expanded into more detailed studies, such as also including exotic Higgs boson decays, additional HNL decay channels, larger simulated samples, the use of timing information, and alternative detector designs.  
The simulation work presented here represents the first step towards a comprehensive evaluation of the experimental potential of FCC-ee in direct searches for BSM. Possible limitations could be solved by innovative experimental solutions that could boost the reach of FCC-ee for other non-standard signals.

\section{Acknowledgements}
The authors would like to thank S. Petcov (SISSA/INFN) for the discussions during the writing of the paper and  for providing comments to the draft. 
R. Gonzalez Suarez is supported by the Swedish Research Council (VR 2017-05092). 
S. Kulkarni is supported by the Austrian Science Fund Elise-Richter grant project number V592-N27.
R. Ruiz acknowledges the support of the Polska Akademia Nauk (grant agreement PAN.BFD.S.BDN. 613. 022. 2021 - PASIFIC 1, POPSICLE). This work has received funding from the European Union's Horizon 2020 research and innovation program under the Sk{\l}odowska-Curie grant agreement No.  847639 and from the Polish Ministry of Education and Science.
A. Thamm was supported by the Australian Research Council through the ARC Discovery Project, DP210101900.

\bibliographystyle{jhep}
\bibliography{references}

\end{document}

%% file: workshopsymbols.tex
\def\beq{\begin{equation}}
\def\eeq#1{\label{#1}\end{equation}}
\def\eeqn{\end{equation}}

\newenvironment{Eqnarray}%
   {\arraycolsep 0.14em\begin{eqnarray}}{\end{eqnarray}}
\def\beqa{\begin{Eqnarray}}
\def\eeqa#1{\label{#1}\end{Eqnarray}}
\def\eeqan{\end{Eqnarray}}

\let\bar=\overbar

\def\lsim{\mathrel{\raise.3ex\hbox{$<$\kern-.75em\lower1ex\hbox{$\sim$}}}}
\def\gsim{\mathrel{\raise.3ex\hbox{$>$\kern-.75em\lower1ex\hbox{$\sim$}}}}

\def\del{\partial}
\def\Dslash{\not{\hbox{\kern-4pt $D$}}}
\def\dslash{\not{\hbox{\kern-2pt $\del$}}}
\def\pslash{\not{\hbox{\kern-2pt $p$}}}
\def\ETmiss{\not{\hbox{\kern-4pt $E$}}_T}

\def\Dlr{\mathrel{\raise1.5ex\hbox{$\leftrightarrow$\kern-1em\lower1.5ex\hbox{$D$}}}}

\def\MSB{{\bar{M \kern -2pt S}}}
\def\msb{{\bar{\scriptsize M \kern -1pt S}}}

\def\drb{{\bar{\scriptsize D \kern -1pt R}}}



%% file: authorlist.tex

\Author{Brigham Young University, Provo, Utah, USA}{C.~B.~Verhaaren}

\Author{CERN, Geneva, Switzerland}{J.~Alimena}

\Author{Durham University, Durham, United Kingdom}{M.~Bauer}

\Author{INFN, Section of Padova, Padova, Italy}{P.~Azzi}

\Author{Institute of Nuclear Physics, Polish Academy of Sciences,
Kracow 31-342, Poland}{R.~Ruiz}

\Author{Johannes Gutenberg University, Mainz, Germany \\ Cornell University, Ithaca, U.S.A.}{M.~Neubert}

\Author{Leiden University, Niels Bohrweg 2, 2333 CA Leiden, The Netherlands}{O.~Mikulenko, M.~Ovchynnikov}

\Author{Universit\'e catholique de Louvain, Louvain-la-Neuve B-1348, Belgium}{M.~Drewes, J.~Klaric}

\Author{University of Geneva, Geneva, Switzerland}{A.~Blondel, C.~Rizzi, A.~Sfyrla, T.~Sharma}

\Author{University of Graz, Graz, Austria}{S.~Kulkarni}

\Author{The University of Melbourne, Victoria 3010, Australia}{A.~Thamm}

\Author{Université Paris-Sorbonne, LPNHE, 4 place Jussieu 75252 Paris France}{A.~Blondel}

\Author{Uppsala University, Uppsala, Sweden}{R.~Gonzalez Suarez, L.~Rygaard}

%% file: theory.tex
\section{Theoretical Landscape}\label{sec:theory}

In this section, the theoretical frameworks considered are briefly summarized.
These representative scenarios are: 
the Phenomenological Type I Seesaw model (Section~\ref{sec:theory_nu}),
axion-like particles (Section~\ref{sec:theory_alps}),
and scalar singlet extensions of the SM (Section~\ref{sec:theory_higgs}).

\subsection{Heavy Neutral Leptons}\label{sec:theory_nu}
\emph{Editors: Marco Drewes, Suchita Kulkarni, Richard Ruiz}

The oscillations between neutrino flavor eigenstates in long-baseline experiments~\cite{Super-Kamiokande:1998kpq,SNO:2002tuh} is one of the most pressing theoretical puzzles in particle physics today. The neutrino masses that produce these oscillations are also interesting because they imply either the existence of new particles and interactions or substantial changes to the SM paradigm~\cite{Feruglio:2015jfa,Ma:1998dn}. 

It is simply not enough to write effective neutrino masses, given the SM's limited particle content and the desire to understand the mechanism (or mechanisms) that render neutrinos so much lighter than charged leptons and quarks. It is also desirable to understand the flavor/mixing pattern among neutrinos and the possible connections to lepton and quark flavors themselves.

Among the most popular solutions to these mysteries are the Seesaw models. These tie the smallness of neutrino masses ($m_{\nu_k}$ with $k=1,2,3$) to the scale of new physics $(\Lambda)$. Depending on the complexity, the scale (or scales) introduced by these models can range from well below the EW scale to the Planck scale. In the most minimal scenarios~\cite{Ma:1998dn}, general arguments only require $\Lambda$ to be below $10^{14}$~GeV. In high-scale Seesaws, light neutrino masses scale inversely with this new physics scale, $m_{\nu_k}\propto 1/\Lambda$.
In low-scale Seesaws, 
the behavior can be more complicated, 
and in some cases
light neutrino masses scale proportionally with this new physics scale, $m_{\nu_k}\propto \Lambda$. 
The manner in which either is implemented can vary widely, cf.~Section~5 in \cite{Agrawal:2021dbo},
and the most minimal, tree-level constructions are known popularly as the Types I~\cite{Minkowski:1977sc,Glashow:1979nm,Gell-Mann:1979vob,Mohapatra:1979ia,Yanagida:1980xy,Schechter:1980gr,Shrock:1980ct}, 
II~\cite{Konetschny:1977bn,Schechter:1980gr,Magg:1980ut,Cheng:1980qt,Lazarides:1980nt,Mohapatra:1980yp}, and III~\cite{Foot:1988aq}
Seesaw models.
Notably, these minimal scenarios are often stepping stones to fuller, more ultraviolet-complete models, including extended gauge theories and grand unified theories. 
Importantly, neutrino mass models predict a plethora of phenomenology that are testable at a variety of low-energy and high-energy experiments~\cite{Atre:2009rg,Deppisch:2015qwa,Cai:2017mow,Agrawal:2021dbo,Bolton:2019pcu}, including particle colliders.

A common feature of several popular solutions to the origin of neutrino masses is the hypothetical existence of heavy, sterile neutrinos $N_i$, or heavy neutral leptons (HNLs) as they are sometimes called. Depending on the precise scenario, they can be Dirac or Majorana fermions, and mediate processes that violate lepton flavor symmetries.
In practice, fermions may be arranged in a way that they form a Majorana state with Dirac-like properties~\cite{Wolfenstein:1981kw,Petcov:1982ya}. This state is known as a pseudo-Dirac fermion and results from underlying symmetries that explain the smallness of neutrino masses~\cite{Shaposhnikov:2006nn,Kersten:2007vk,Ibarra:2010xw,Moffat:2017feq}. This leads to a phenomenology that practically interpolates between these limiting cases in the sense that the ratio between the rates of the lepton number violating and conserving decays $(R_{ll})$ can smoothly interpolate between $R_{ll}=0$ and $R_{ll}=1$~\cite{Anamiati:2016uxp}.
Searches for heavy Dirac and Majorana neutrinos at $e^+e^-$ facilities have a long history~\cite{Tsai:1971vv,Gronau:1984ct,Petcov:1984nf,Dittmar:1989yg}, and if they are discovered at the LHC, FCC-ee would be a natural program to study their properties~\cite{Blondel:2014bra,Duarte:2015iba,PhysRevD.92.075002,Duarte:2016miz,Antusch:2016vyf,Cai:2017mow,Antusch:2017pkq,Duarte:2018kiv,Ding:2019tqq,Barducci:2020icf,Blondel:2021mss,Zapata:2022qwo,Shen:2022ffi,Barducci:2022hll}. 

If HNLs mix with the SM neutrinos, they can participate in the SM weak interaction via the couplings 
 \begin{subequations}
 \label{eq:lag_hnl}
\begin{align}
 \mathcal{L}^{\rm Int}_{\rm Type~I} &= \mathcal{L}^{W} + \mathcal{L}^{Z} + \mathcal{L}^{H}, \quad\text{where}
 \\
 \mathcal{L}^{W} &= -\frac{g_W}{\sqrt{2}}\sum_{\ell=e}^\tau \sum_{i=1}^{n_s} \overline{N_{i}} V^*_{\ell i} W^+_\mu \gamma^\mu P_L \ell^- + {\rm H.c.},
 \\
 \mathcal{L}^{Z} &= -\frac{g_W}{2\cos\theta_W}\sum_{\ell=e}^\tau \sum_{i=1}^{n_s} \overline{N_{i}} V^*_{\ell i} Z_\mu \gamma^\mu P_L \nu_\ell + {\rm H.c.},
 \\
 \mathcal{L}^{H} &= -\frac{g_W}{2M_W} h \sum_{\ell=e}^\tau \sum_{i=1}^{n_s} \overline{N_{i}} V^*_{\ell i} m_{N_{i}} P_L \nu_\ell + {\rm H.c.}
 \end{align}
 \end{subequations} 
Here, $N_1, \ldots N_{n_s}$ are the heavy mass eigenstates of the theory. 
This model \eqref{eq:lag_hnl} is a common HNL benchmark for the pure type I Seesaw we use in this study.
In extended models, the HNLs may have extra interactions, such as new gauge interactions~\cite{Pati:1974yy,Mohapatra:1974hk,Mohapatra:1974gc,Senjanovic:1975rk,Senjanovic:1978ev,Langacker:1980js,Hewett:1988xc,Faraggi:1990ita,Buchmuller:1991ce}.
The number of right-handed neutrino chiral eigenstates $n_s$ is not constrained by gauge anomaly considerations in the model~\eqref{eq:lag_hnl}
because the chiral states are gauge singlets.
Here, the $V_{\ell N_i}$ are the complex-valued, active-sterile mixing matrix elements and describe the coupling  between the heavy mass eigenstate $i$ and lepton flavor state $\ell$.

The Lagrangian \eqref{eq:lag_hnl} approximates interactions to first order in the parameter $\vert V_{\ell i}\vert$. 
In this phenomenological framework, the masses of $N_i$ $(m_{N_i})$ and $V_{\ell i}$ are taken to be independent. Hypothesizing connections to other physics, e.g., the relic abundance of DM or the matter-antimatter asymmetry of the observable universe, can greatly constrain masses and mixing, cf.~Section~\ref{CosmoConnection}. For simplicity, the analysis of Section~\ref{sec:analysis_hnl} considers only the lightest heavy mass eigenstate $N_1$, denoted by $N$, with mass and mixing $m_N$ and $V_{\ell N}$. 
It is important to stress that considering only one HNL is for bench-marking and discovery purposes; realistic scenarios usually contain multiple mass eigenstates.

In analogy to the SM effective field theory (SMEFT) framework~\cite{Grzadkowski:2010es,Brivio:2017vri,Gauld:2015lmb}, the above Lagrangian can be systematically extended by higher dimensional operators in a framework known as $\nu$SMEFT~\cite{delAguila:2008ir,Aparici:2009fh,Elgaard-Clausen:2017xkq}, which can parameterize ultraviolet completions.

\begin{figure}[t!]
\begin{center}
\includegraphics[width=0.45\textwidth]{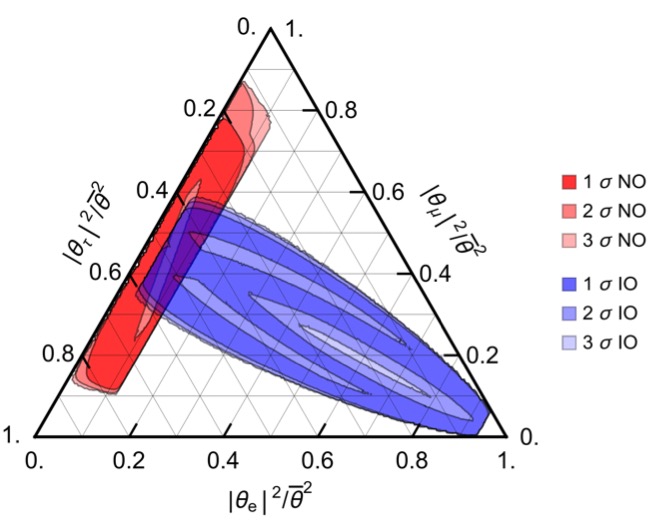}\includegraphics[width=0.5\textwidth]{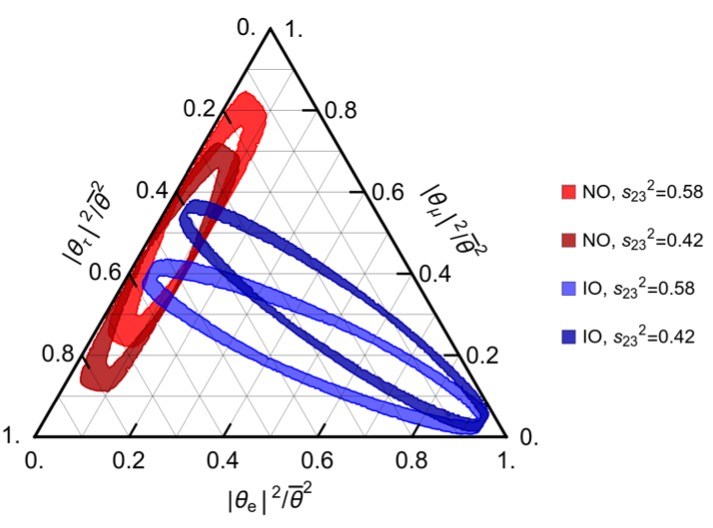}
\caption{
Allowed range for the relative magnitude of the HNL couplings to individual SM flavors in the model \eqref{eq:hnl_mee_def} with $n_s=2$, plot taken from~\cite{Abdullahi:2022jlv}.  
 \textit{Left panel}: The range of relative flavor mixings
 $(\sum_i|V_{\ell i}|^2)/(\sum_{i,\ell}|V_{\ell i}|^2)$
 consistent with the current neutrino oscillation data, cf.~e.g.~\cite{Drewes:2016jae,Caputo:2017pit,Drewes:2018gkc}. The  contours correspond to the allowed $\Delta \chi^2$ range taken from~\cite{Esteban:2020cvm} for the case of normal (red) and inverted (blue) light neutrino mass ordering.
\textit{Right panel}: The projected $90\%$ CL contours for the relative mixings after 14 years of data taking at DUNE~\cite{DUNE:2020jqi}, assuming maximal CP violation $\delta=-\pi/2$ and two benchmark values of the PMNS angle $\theta_{23}$, taken from the DUNE TDR~\cite{DUNE:2020ypp}, as indicated in the legend.
FCC-ee can measure these ratios to the percent level in displaced HNL decays \cite{Antusch:2017pkq}.
}
\label{fig:triangle}
\end{center}
\end{figure}

In the minimal Type I Seesaw model, where $n_s=2$,
the requirement to reproduce the observed pattern of light neutrino masses and mixing imposes testable constraints on the relative size of the HNL couplings $|V_{\ell i}|^2$ to individual SM flavors~\cite{Hernandez:2016kel,Drewes:2016jae,Drewes:2018gkc,Chrzaszcz:2019inj}. 
These will improve in the future with the Deep Underground Neutrino Experiment (DUNE)~\cite{DUNE:2020ypp}, cf.~Figure~\ref{fig:triangle}, leading to a prediction that can be tested with FCC-ee.
The position in the triangle in Figure~\ref{fig:triangle} is entirely determined by the low energy phases in the Pontecorvo-Maki-Nakagawa-Sakata (PMNS) matrix. 
The number of events observed in displaced decays of HNLs produced during the $Z$ pole run permits the determination of the relative mixing
$|V_{\ell i}|^2/(\sum_\ell |V_{\ell i}|^2)$
at the percent level~\cite{Antusch:2017pkq}, allowing for the Majorana phase in the light neutrino mixing matrix to be indirectly constrained ~\cite{Drewes:2016jae,Caputo:2016ojx}.
For $n_s=3$, the model is less constrained, and making a testable prediction would require an independent determination of the lightest neutrino mass in the SM, cf.~Figure~11 in~\cite{Chrzaszcz:2019inj}. Beyond minimal models, measuring the $|V_{\ell i}|^2/(\sum_\ell |V_{\ell i}|^2)$ can give insight into flavor at charge-parity (CP) symmetries of the neutrino mixing matrix
~\cite{King:2013eh,Feruglio:2019ybq,Xing:2020ijf}, providing a hint towards possible ultraviolet completions.

\subsubsection{Phenomenology of Dirac and Majorana Heavy Neutral Leptons}\label{sec:theory_nu_pheno}

\begin{figure*}
\centering
\includegraphics[width=1\textwidth]{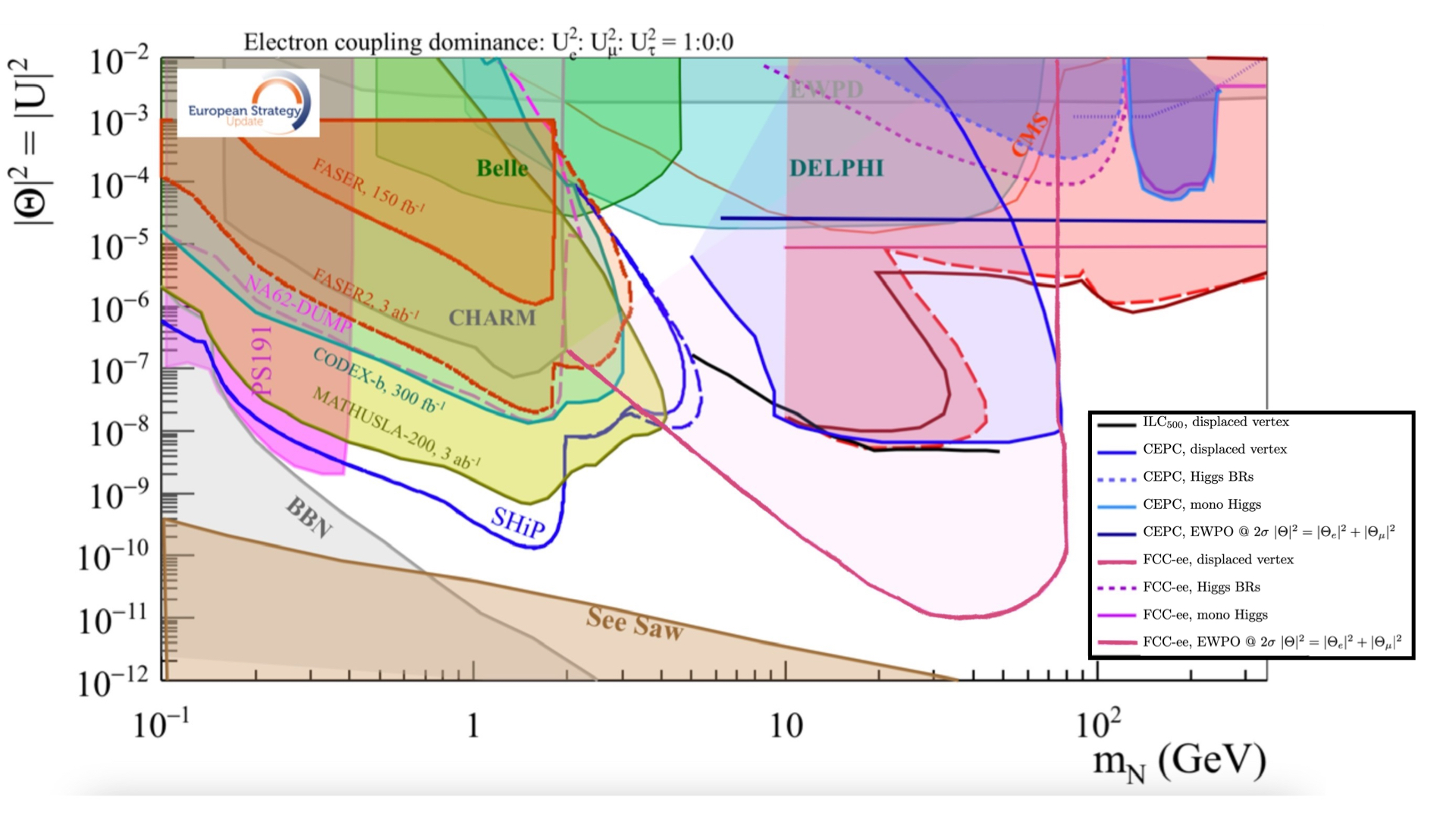}
\caption{90\% CL exclusion limits for a Heavy Neutral Lepton mixed with the electron neutrino, as presented in the European Strategy for Particle Physics Briefing Book~\cite{EuropeanStrategyforParticlePhysicsPreparatoryGroup:2019qin}. The FCC-ee curves are in (overlined) dark purple---for FCC-ee, this is equivalent to a plot as function of the sum of matrix elements squared $\rm {|U_N|^2}$. The curve below the $Z$ boson mass corresponds to the combined LLP and prompt analysis performed with $10^{12} Z$ in Ref.~\cite{Antusch:2016ejd}. The horizontal limit at high masses results from the effect of light-heavy neutrino mixing on the EW precision observables and remains valid up to O(1000 TeV).}
\label{fig:HNLsummaryBriefing}
\end{figure*}

\begin{figure*}
\centering
\includegraphics[width=0.9\textwidth]{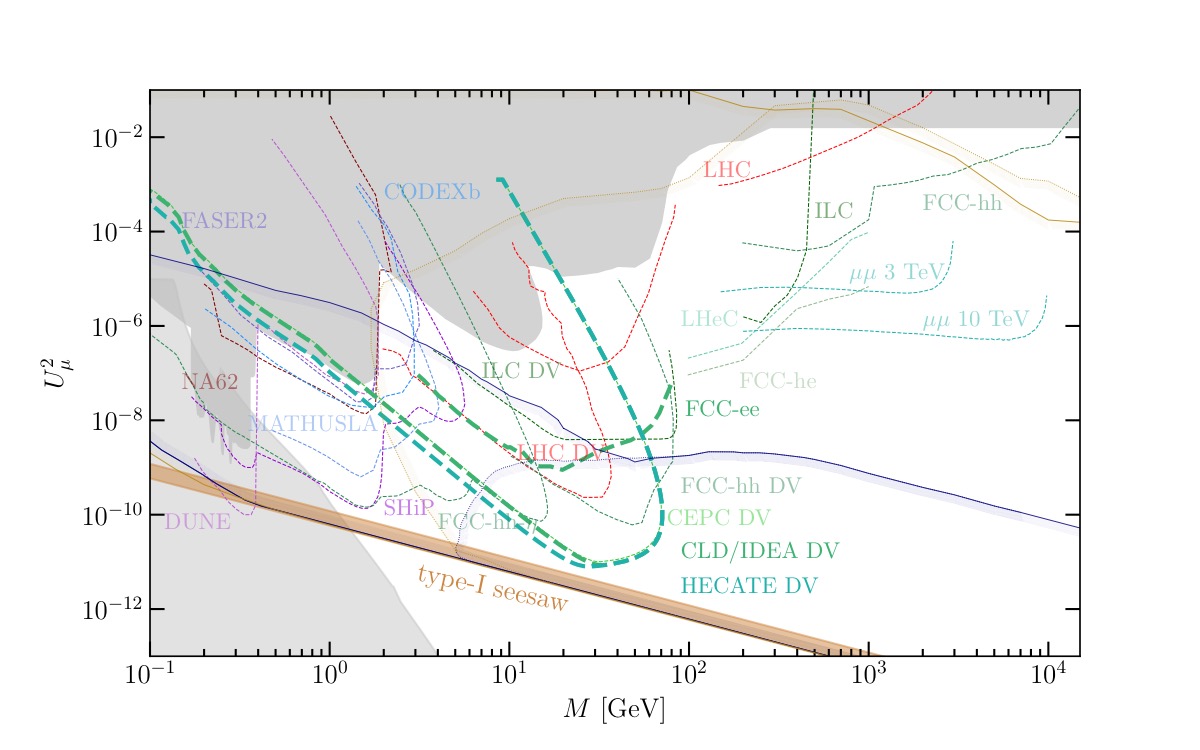}
\caption{
\emph{Bold green line:} Sensitivity 
of displaced vertex searches at FCC-ee. The parameter region inside the curves
corresponds to more than four observed HNL decays 
with $|V_{ \ell N }|^2=\delta_{\ell \mu}U_\mu^2$
from 
$5\times 10^{12}$ $Z$ bosons, assuming no background events and $95\%$ reconstructed HNL decays (i.e., all decays except the invisible decay) inside the main detectors based on the IDEA or CLD design with a displacement of over 400~$\mu$m. Based on Tables 7.2 and 7.3 in \cite{FCC-CDR2} with 1~m of instrumentation required for detection, we assume a cylinder of length $l = 8.6$~m and radius $r = 5$~m (CLD) or $l = 11$~m and  $r = 4.5$~m (IDEA) as fiducial volumes. The resulting curves for the CLD and IDEA detectors are visually indistinguishable.
For comparison, we show what CEPC can achieve  with  $4.2\times 10^{12}$ $Z$ bosons \cite{CEPCtalk} for an IDEA-type detector \cite{CEPCStudyGroup:2018ghi}.
\emph{Bold turquoise line:} Gain in sensitivity if the maximal observable displacement is increased with HECATE-like detectors~\cite{Chrzaszcz:2020emg} with $l = 60$~m, $r = 15$~m at two IPs.
\emph{Medium gray:} Constraints on the mixing of HNLs from past experiments~\cite{CHARM:1985nku,Abela:1981nf,Yamazaki:1984sj,E949:2014gsn,Bernardi:1987ek,NuTeV:1999kej,Vaitaitis:2000vc,CMS:2018iaf,DELPHI:1996qcc,ATLAS:2019kpx,CMS:2022fut}.
 \emph{Colorful lines:}
Estimated sensitivities of the main HL-LHC detectors 
\cite{Izaguirre:2015pga,Drewes:2019fou,Pascoli:2018heg} and NA62 \cite{Drewes:2018gkc},
compared to the sensitivities of selected planned or proposed experiments (DUNE~\cite{Ballett:2019bgd}, FASER2~\cite{FASER:2018eoc}, SHiP~\cite{SHiP:2018xqw,Gorbunov:2020rjx}, MATHUSLA~\cite{Curtin:2018mvb}, CODEX-b~\cite{Aielli:2019ivi}, cf.~\cite{Agrawal:2021dbo} for a more complete list), prompt searches at FCC-ee or CEPC \cite{Shen:2022ffi,BayNielsen:2017yws},
and searches at selected other proposed future colliders 
(FCC-hh~\cite{Antusch:2016ejd,Antusch:2018bgr,Pascoli:2018heg,Boyarsky:2022epg},
ILC \cite{Mekala:2022cmm,Antusch:2016vyf}
LHeC and FCC-he \cite{Antusch:2019eiz},
and muon colliders \cite{ZhenLiuTeamHNLMuC}, with DV indicating displaced vertex searches).
The curves from \cite{Izaguirre:2015pga,Shen:2022ffi,BayNielsen:2017yws} were re-scaled for a consistent integrated luminosity with \cite{Drewes:2019fou,Pascoli:2018heg}.
The sensitivity of FCC-ee and other future colliders can be further improved with dedicated long-lived particle detectors \cite{Wang:2019xvx,Chrzaszcz:2020emg,Giffin:2022rei,Boyarsky:2022epg}.
\emph{Brown band:} Indicative lower bound on the total HNL mixing $U_e^2+U_\mu^2+U_\tau^2$ from the requirement to explain the light neutrino oscillation data \cite{Esteban:2020cvm}.
The band width corresponds to varying the light neutrino mass ordering and the lightest neutrino mass.
 The matter-antimatter asymmetry of the universe \cite{Canetti:2012zc} can be explained by low scale leptogenesis  \cite{Akhmedov:1998qx,Asaka:2005pn,Pilaftsis:2003gt} together with the light neutrino properties inside the mustard (violet) hashed contours with three \cite{Drewes:2021nqr} (two \cite{Klaric:2021cpi}) HNL flavours;  solid and dashed contours indicate vanishing and thermal initial conditions in the early universe, respectively.
\emph{Light gray:} Lower bound on $U_\mu^2$
 from BBN~\cite{Sabti:2020yrt,Boyarsky:2020dzc}.
Plot adapted from \cite{Abdullahi:2022jlv}.
}
\label{fig:HNLsummary}
\end{figure*}

In the kinematically accessible regime, FCC-ee is an excellent machine to discover HNLs~\cite{Blondel:2014bra} and study their properties. An analysis was completed for prompt HNL signals~\cite{Antusch:2016ejd} and reproduced in Ref.~\cite{FCC-CDR2}. A comparison for various machines and setups was compiled for the European Strategy for Particle Physics Briefing Book~\cite{EuropeanStrategyforParticlePhysicsPreparatoryGroup:2019qin}\footnote{Figure 8.19}. The sensitivity to active-sterile mixing, labeled here by $\Theta$, is shown in the summary figure, Figure~\ref{fig:HNLsummaryBriefing}. Figure~\ref{fig:HNLsummary} shows an updated estimation of different sensitivities for current and proposed detectors including an FCC-ee displaced vertex analysis. 

\begin{figure*}
\centering
\includegraphics[width=0.9\textwidth]{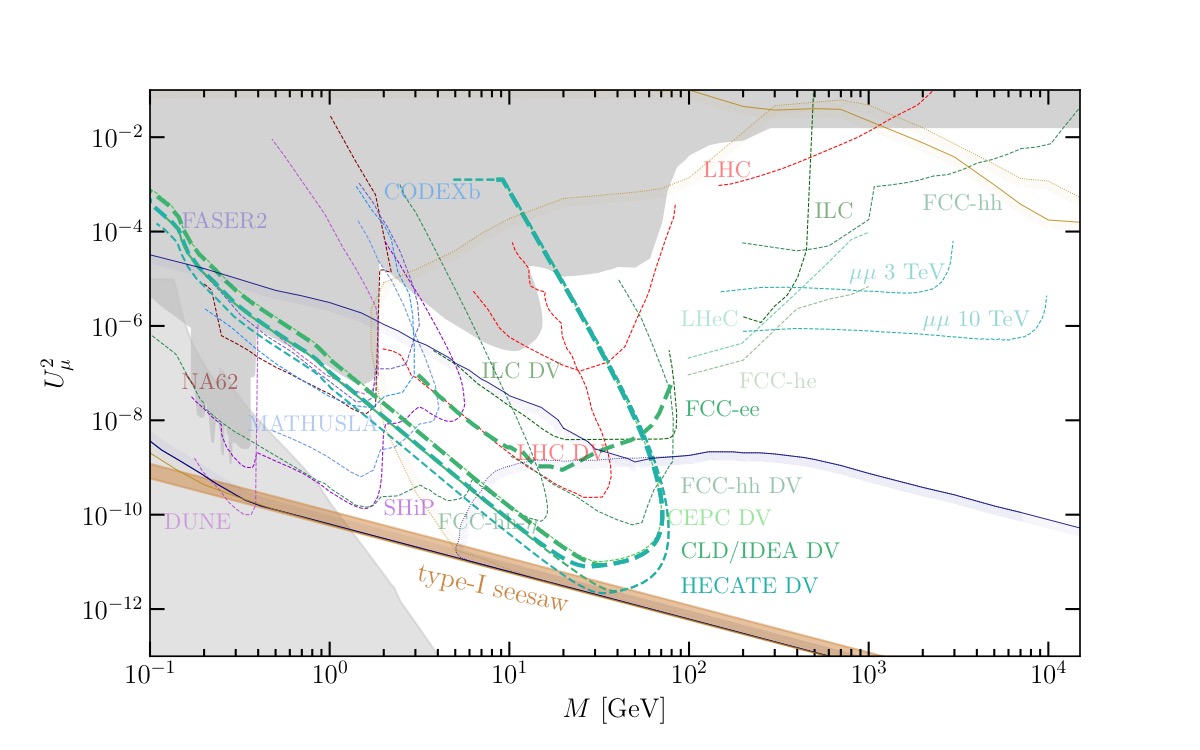}
\caption{
Comparison of the parameter regions in which four events (bold lines) and one event (non-bold lines) are expected in the IDEA/CLD detector or HECATE, with the same conventions and assumptions as in Figure~\ref{fig:HNLsummary}.
}
\label{fig:HNLsummary_1event}
\end{figure*}

Figure~\ref{fig:HNLsummary_1event} shows the four and one event contours. The four event contour corresponds to the 95\% confidence level (CL) exclusion in the absence of signal events. The one event contour shows that for the analysis of the LLP signatures, there still is a 63\% probability to observe one event, which, in the absence of background events, could be sufficient for discovery, all the way down to the see-saw limit around 20-40 GeV.

If $N$ is a Majorana fermion, then it can mediate processes that conserve lepton number as well as those that violate lepton number. Likewise, if lepton number is violated, then neutrinos must in principle possess  Majorana properties~\cite{Schechter:1981bd,Hirsch:2006yk}, though the amount of lepton number violation (LNV) in practice depends on the underlying model~\cite{Hirsch:2006yk,Duerr:2011zd}.
Dirac neutrinos can only participate in processes that conserve lepton number.
Therefore, differences between Dirac and Majorana $N$ are closely related to differences between lepton number conservation (LNC) and LNV.
The availability of LNV decay modes, for instance, leads to a Majorana $N$  having a width $(\Gamma_N)$ that is twice as large as a Dirac $N$. This implies that a Majorana $N$ has a mean lifetime $(\tau_N)$, or mean displacement $(d_N)$, that is half as long as that of a Dirac $N$.

Section~\ref{sec:md} assumes a simple phenomenological model \eqref{eq:lag_hnl} with $n_s=1$, i.e., only one mass eigenstate $N$ that is either a Dirac or a Majorana fermion. Though this phenomenological model cannot completely reproduce the light neutrino oscillation data, it is sufficient to capture the collider phenomenology of the pure Dirac and Majorana HNLs benchmarks experimentally. Nature may be somewhere in the middle of these two cases.
Therefore, observables that quantify differences between LNV and LNC processes, such as $R_{ll}$ as defined in~\cite{Anamiati:2016uxp} or $\mathcal{A}$ as defined in~\cite{Bray:2007ru,Ruiz:2020cjx}, can be used to interpolate between the benchmarks.

There is a rich phenomenology that connects $R_{ll}$, $\mathcal{A}$, and the decay rates of HNLs into different SM flavors to the mechanism that generates light neutrino masses. This connection can be accurately probed by studying the HNL properties at FCC~\cite{Petcov:1984nf,Blondel:2021mss,deGouvea:2021ual} or other  colliders \cite{Bray:2007ru,Anamiati:2016uxp,Hernandez:2018cgc,Ruiz:2020cjx}. 

\FloatBarrier
\subsubsection{Probing Cosmological Questions}\label{CosmoConnection}

In addition to generating light neutrino masses, HNLs can, depending on their mass, affect the history of the universe in different ways~\cite{Asaka:2005an,Asaka:2005pn,Drewes:2013gca}.
From a cosmological viewpoint, the most important motivation for HNL searches may be their potential connection to the origins of baryonic matter and DM.

\paragraph{Leptogenesis:} 
Leptogenesis~\cite{Fukugita:1986hr} provides an explanation for the matter-antimatter asymmetry in the observable universe~\cite{Canetti:2012zc} and relates it to the properties of neutrinos.
In the most popular scenario (based on the Type I Seesaw), HNLs generate the BAU via their CP-violating interactions with the thermal plasma. While it was originally believed that this mechanism only operates for HNL masses far above the EW scale~\cite{Davidson:2002qv}, it is now established that sub-TeV HNLs can produce the observed BAU during their production~\cite{Akhmedov:1998qx,Asaka:2005an,Asaka:2005pn} or freeze-out and decay~\cite{Pilaftsis:2003gt,Pilaftsis:2005rv}. This implies that direct experimental searches have the potential to probe the origin of matter~\cite{Chun:2017spz}. 
If any HNLs with masses at or below the EW scale are discovered in the near future, FCC-ee would provide a powerful tool to study their properties and test their connection to the BAU. 
A simple construction that supports low-scale leptogenesis is the Neutrino Minimal Standard Model ($\nu$MSM) \cite{Asaka:2005an}. Here, two HNLs simultaneously explain the neutrino masses and the BAU \cite{Asaka:2005pn} for a wide range of experimentally accessible masses, cf.~Figure~\ref{fig:HNLsummary}. Due to its minimality, the model is highly testable \cite{Hernandez:2016kel,Drewes:2016jae}. In particular, leptogenesis constrains the flavor mixing pattern beyond the experimental fits shown in Figure~\ref{fig:triangle}, which can be tested by comparing flavored branching ratios in displaced decays. Finally, if accessible, HNL oscillations in the detectors are sensitive to the HNL mass splitting \cite{Antusch:2017pkq}, which is a crucial parameter for leptogenesis.

\paragraph{Dark matter:} HNLs with sufficiently small masses and mixing angles could be viable DM candidates \cite{Dodelson:1993je}. Constraints on the HNL lifetime and from indirect searches restrict the range of masses and mixings to values that are inaccessible to direct searches at colliders, cf.~\cite{Drewes:2016upu,Boyarsky:2018tvu} for reviews. 
However, FCC-ee can indirectly probe sterile neutrino SM scenarios by searching for signatures of other particles that were involved in the DM production.

HNLs can be resonantly produced in the early universe through their mixing-suppressed weak interactions if the lepton asymmetry at temperatures around the quantum chromodynamics (QCD) crossover greatly exceeded the BAU~\cite{Shi:1998km,Laine:2008pg,Ghiglieri:2015jua,Venumadhav:2015pla}. In the $\nu$MSM, this large lepton asymmetry can be generated by heavier HNLs that are also responsible for the BAU and neutrino masses~\cite{Shaposhnikov:2008pf}. The first parameter space studies~\cite{Canetti:2012vf,Canetti:2012kh,Ghiglieri:2020ulj} suggest that this is possible only for comparably small mixing angles, possibly making FCC-ee or a similar machine the only facility at which these HNLs could be discovered.
If the HNLs have additional gauge interactions (cf.~e.g.~\cite{Kusenko:2010ik,Li:2010rb,Escudero:2016tzx,Bezrukov:2009th,Bezrukov:2012as,Nemevsek:2012cd}), the extended gauge sector can be probed directly or indirectly at FCC-ee. If the DM is produced via the decay of a singlet~\cite{Shaposhnikov:2006xi,Petraki:2007gq,Boyanovsky:2008nc} or charged~\cite{Frigerio:2014ifa,Drewes:2015eoa} scalar during freeze-out or freeze-in~\cite{Merle:2013wta,Konig:2016dzg}, precision studies of the SM Higgs and of the portal can shed light on the mechanism. Most of these possibilities have not been studied in detail to date.

\FloatBarrier

\subsection{Axion-Like Particles}\label{sec:theory_alps}
\emph{Editors: Martin Bauer, Matthias Neubert, Andrea Thamm} 

Many models that address open, fundamental problems of the SM are governed by global symmetries. If an approximate global symmetry is spontaneously broken, a pseudo Nambu-Goldstone boson appears in the theory that is light compared to the symmetry breaking scale. If this pseudo Nambu-Goldstone boson is a pseudoscalar, it is often referred to as an axion-like particle or ALP. The ALP's lightness singles it out as a uniquely promising experimental target that could open a first window onto high-scale new physics beyond the SM.

ALPs appear in many models that address open, fundamental  problems in the SM. The most prominent example is the QCD axion, which was introduced in the 1980s to solve the strong CP problem~\cite{Peccei:1977hh,Peccei:1977ur,Weinberg:1977ma,Wilczek:1977pj} and found to simultaneously account for the observed DM relic abundance~\cite{Preskill:1982cy,Dine:1982ah}. QCD axions are typically very light, and these models are plagued by the ``axion quality'' problem, in which quantum gravity corrections destabilize the minimum of the axion potential, thereby reintroducing the strong CP problem~\cite{Kamionkowski:1992mf,Barr:1992qq,Holman:1992us,Ghigna:1992iv}. Heavy-axion solutions to the strong CP problem circumvent this issue and so motivate ALPs with MeV-to-TeV scale masses~\cite{Rubakov:1997vp,Holdom:1982ex,Berezhiani:2000gh,Hook:2014cda,Fukuda:2015ana,Gherghetta:2016fhp,Dimopoulos:2016lvn,Agrawal:2017ksf,Gaillard:2018xgk}. ALPs in this mass range could also result from the breaking of global symmetries in low scale supersymmetric~\cite{Nelson:1993nf,Goh:2008xz,Bellazzini:2017neg} or composite Higgs models \cite{Gripaios:2009pe,Ferretti:2013kya,Belyaev:2015hgo,Ferretti:2016upr}.  Phenomenologically, they can also lead to successful EW baryogenesis~\cite{Jeong:2018ucz}.

An ALP dominantly couples to SM particles via dimension-5 operators,
\begin{equation}\label{LALP}
\begin{aligned}
   {\cal L}_{\rm eff}
   &= \frac12 \left( \partial_\mu a\right)\!\left( \partial^\mu a\right) - \frac{m_{a,0}^2}{2}\,a^2
    +  \sum_\psi \frac{c_{ff}}{2}\,\frac{\partial^\mu a}{f}\,\bar \psi\gamma_\mu\gamma_5 \psi \\
    &\quad\mbox{}+ c_{GG}\,\frac{\alpha_s}{4\pi}\,\frac{a}{f}\,G_{\mu\nu}^a\,\tilde G^{\mu\nu,a}
    + c_{\gamma\gamma}\,\frac{\alpha}{4\pi}\,\frac{a}{f}\,F_{\mu\nu}\,\tilde F^{\mu\nu} \\
  &\quad\mbox{}+ c_{\gamma Z}\,\frac{\alpha}{2\pi s_w\space c_w}\,
    \frac{a}{f}\,F_{\mu\nu}\,\tilde Z^{\mu\nu}
    + c_{ZZ}\,\frac{\alpha}{4\pi s_w^2\space c_w^2}\,\frac{a}{f}\,Z_{\mu\nu}\,\tilde Z^{\mu\nu} 
    + c_{WW}\,\frac{\alpha}{2\pi s_w^2}\,\frac{a}{f}\,W_{\mu\nu}^+\,\tilde W^{-\mu\nu} \,,
\end{aligned}
\end{equation}
where $G_{\mu\nu}^a$ is the field-strength tensor of $SU(3)_c$, while $F_{\mu\nu}$, $Z_{\mu\nu}$ and $W_{\mu\nu}^+$ describe the photon, $Z$, and $W$ boson in the broken phase of EW symmetry. The dual field-strength tensors are denoted by $\tilde F^{\mu\nu}=\frac12\epsilon^{\mu\nu\alpha\beta} F_{\alpha\beta}$, etc.\ (with $\epsilon^{0123}=1$); $\alpha_s$ and $\alpha$ are the QCD coupling and fine-structure constants, respectively; $s_w$ and $c_w$ denote the sine and cosine of the weak mixing angle; and the sum runs over all fermion mass eigenstates $\psi$. The suppression scale $f$ is related to the new physics scale $\Lambda$ via $\Lambda = 4\pi f$, and to the axion decay constant $f_a$ by $f_a = -f /(2c_{GG})$. The ALP dominantly interacts with the Higgs boson via dimension-6 and -7 operators,
\begin{equation}
   {\cal L}_{\rm eff}^H
   = \frac{c_{ah}}{f^2} \left( \partial_\mu a\right)\!\left( \partial^\mu a\right) H^\dagger H
    + \frac{c_{Zh}}{f^3} \left( \partial^\mu a\right) 
    \left( H^\dagger\,iD_\mu\, H + \mbox{h.c.} \right) H^\dagger H \,.
\end{equation}

\begin{figure}[hbtp]
\centering
\includegraphics[width=0.75\textwidth]{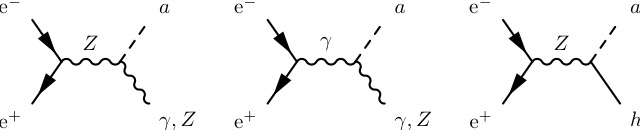}
\caption{ALP production processes in electron-positron collisions.}
\label{fig:ALPsFeynmanDiagram}
\end{figure}

At FCC-ee, ALPs are predominantly produced in association with a photon, $Z$ boson, or Higgs boson, as shown in the Feynman diagrams in Figure~\ref{fig:ALPsFeynmanDiagram}, or via exotic $Z$ and Higgs decays. Resonant production of an ALP, e.g., $e^+ e^- \to a$, is possible but suppressed by $m_e^2/(4\pi f)^2$. ALP production in vector boson fusion has been considered in Ref.~\cite{Yue:2021iiu} and detection prospects in light-by-light scattering in Ref.~\cite{Zhang:2021sio,Coelho:2020saz}.  

The differential cross sections for associated $\gamma a/Za/ha$ production are given by~\cite{Bauer:2017ris,Bauer:2018uxu}
\begin{align} \label{eq:ALPdiffcross1}
   \frac{d \sigma(e^+ e^-\to\gamma a)}{d\Omega} =
   & \, \frac{\alpha \alpha^2(s)}{128 \pi^3} \frac{s^2}{f^2} \left(1 - \frac{m_a^2}{s}\right)^3 \left(1 + \cos^2 \theta \right) \left(|V_\gamma(s)|^2 + |A_\gamma(s)|^2\right), \\ \vspace{0.1cm} \label{eq:ALPdiffcross2}
     \frac{d \sigma(e^+ e^-\to Z a)}{d \Omega} =
    & \, \frac{\alpha \alpha^2(s)}{128 \pi^3} \frac{s^2}{f^2} \, \lambda^{\frac{3}{2}} \left(x_a, x_Z \right)\left(1 + \cos^2 \theta \right) \left(|V_Z(s)|^2 
+ |A_Z(s)|^2\right), \\ \vspace{0.1cm} \label{eq:ALPdiffcross3}
      \frac{d \sigma(e^+ e^-\to h a)}{d\Omega} 
   = & \, \frac{2\pi^3 \alpha}{c_w^2 s_w^2} \frac{|c_{Zh}|^2}{f^2}  \frac{s \, m_Z^2}{(s-m_Z^2 )^2}\, \lambda^{\frac{3}{2}} \left(x_a,x_h \right) \sin^2 \theta \, (g_V^2 + g_A^2) \,,
\end{align}
where $\lambda(x,y)=(1-x-y)^2-4xy$, $x_i =(m_i^2/s)$, $\sqrt{s}$ is the center-of-mass energy, and $\theta$ describes the scattering angle of the photon, $Z$, or Higgs boson relative to the beam axis. The vector and axial-vector form factors are given by
\begin{align}\label{eq:ALPdiffcross4}
V_\gamma(s) &= \frac{c_{\gamma \gamma}}{s} + \frac{g_V}{2 c_w^2 s_w^2}\frac{c_{\gamma Z}}{s - m_Z^2+ i m_Z \Gamma_Z}\, ,
\qquad
A_\gamma(s) = 
\frac{g_A}{2 c_w^2 s_w^2}\frac{c_{\gamma Z}}{s - m_Z^2+ i m_Z \Gamma_Z} \,, \\ \vspace{0.1cm}
V_Z(s) &= \frac{1}{c_w s_w}\frac{c_{\gamma Z}}{s} + \frac{g_V}{2 c_w^3 s_w^3}\frac{c_{Z Z}}{s - m_Z^2+ i m_Z \Gamma_Z} \,,  
\qquad
A_Z(s)= \frac{g_A}{2 
c_w^3 s_w^3}\frac{c_{Z Z}}{s - m_Z^2+ i m_Z \Gamma_Z}\,,
\end{align}
with $g_V = 2 s_w^2 - 1/2$, $g_A=-1/2$, and $\Gamma_Z$ is the total width of the $Z$ boson. The process where an ALP is radiated off an initial-state electron exhibits an additional suppression of $(m_e^2/s)$.

The integrated cross section of $e^+ e^-\to\gamma a$ below the $Z$ pole is dominated by the photon contribution, which is proportional to $c_{\gamma \gamma}$, while above the $Z$ pole the process proportional to $c_{\gamma Z}$ also contributes. Combining these measurements at low and high energies therefore enables us to access these couplings separately. At the $Z$ pole, the cross section becomes
\begin{equation}
   \sigma(e^+ e^-\to\gamma a)
   \approx \frac{\alpha}{24 \pi^2}\,\alpha^2(m_Z^2) \left( 1 - \frac{m_a^2}{m_Z^2} \right)^3       
    \left[ \frac{|c_{\gamma\gamma}|^2}{f^2} 
    + \frac{m_Z^2}{\Gamma_Z^2}\,\frac{|c_{\gamma Z}|^2}{16 s_w^4 c_w^4\,f^2} \right] .
\end{equation}
The contribution from the $Z$ boson propagator is enhanced by $(m_Z^2/\Gamma_Z^2) \sim 1336$, which allows one to directly access the coupling $c_{\gamma Z}$ (as long as $c_{\gamma \gamma}$ is not much bigger than $c_{\gamma Z}$).
ALPs can also be produced in exotic decays of $Z$ and Higgs bosons \cite{Bauer:2017ris,Bauer:2018uxu,Cacciapaglia:2021agf}. The exotic decay rates are given by
\begin{align}
\Gamma(Z\to \gamma a)&=\frac{\alpha\,\alpha(m_Z)\,m_Z^3}{96 \pi^3 s_w^2 c_w^2 f^2}\,\big| c_{\gamma Z} \big|^2
    \left( 1 - \frac{m_a^2}{m_Z^2} \right)^3 \,,\\
\Gamma(h \to Za)&=\frac{m_h^3 v^2}{64\pi\,f^6}|c_{Zh}|^2\lambda^{3/2}\Big(\frac{m_Z^2}{m_h^2},\frac{m_a^2}{m_h^2}\Big)\, \label{eq:widthHZa},\\
\Gamma(h \to aa)&=\frac{m_h^3\,v^2}{32\pi\,f^4}|c_{ah}|^2\left(1-\frac{2m_a^2}{m_h^2}\right)^2\sqrt{1-\frac{4m_a^2}{m_h^2}}\,.\label{eq:widthHaa}
\end{align}

\begin{figure}[t!]
\centering
\includegraphics[width=0.25\textwidth]{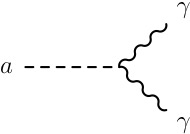} \hspace{3cm}
\includegraphics[width=0.25\textwidth]{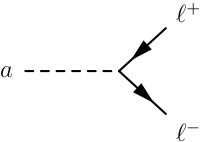}
\caption{ALP decay processes at FCC-ee.}
\label{fig:ALPsDecayFeynmanDiagram}
\end{figure}

Once produced, ALPs lead to a variety of signatures inside the detector. Very long-lived ALPs, for example, escape the detector and lead to a signature with missing momentum. ALPs with somewhat shorter lifetimes may decay into gauge bosons, leptons, and quarks inside the detector. The photon and lepton decay channels are shown in Figure~\ref{fig:ALPsDecayFeynmanDiagram}. Their corresponding decay widths are given by
\begin{align}
   \Gamma(a\to\gamma\gamma)
   &= \frac{\alpha^2\space m_a^3}{64\pi^3 f^2}  c_{\gamma\gamma}^2 \,, \\
   \Gamma(a\to\ell^+\ell^-)
    &= \frac{m_a\space m_\ell^2}{8\pi f^2}\,c_{\ell\ell}^2\,\sqrt{1-\frac{4m_\ell^2}{m_a^2}} \,.
\end{align}
An ALP decay into hadrons can be computed perturbatively for relatively large ALP masses, i.e., $m_a \gg \Lambda_{\text{QCD}}$. The decay width into bottom quarks specifically is given by
\begin{equation}
   \Gamma(a\to b\bar b) = \frac{3 \, m_a\space m_b^2(m_a)}{8\pi f^2}\,
    c_{bb}^2\,\sqrt{1-\frac{4m_b^2}{m_a^2}} \,,
\end{equation}
and similarly for $\Gamma(a\to c\bar c)$. The decay rate into light quarks ($u$, $d$, $s$) can be computed using quark-hadron duality and is given by~\cite{Spira:1995rr,Bauer:2017ris}
\begin{equation}
   \Gamma(a\to\mbox{light hadrons}) 
   = \frac{\alpha_s^2(m_a)\space m_a^3}{8\pi^3 f^2}   
    \left[ 1 + \frac{83}{4}\,\frac{\alpha_s(m_a)}{\pi} \right] 
    \left| C_{GG}^\text{eff}(m_a)\right|^2 ,
\end{equation}
where the ALP couplings to both gluons and quarks contribute via
\begin{equation}
C_{GG}^\text{eff}(m_a) = c_{GG} + \frac12\space\sum_{q\ne t}\space
      c_{qq} \,B_1\left(\frac{4m_q^2}{m_a^2}\right) \,.
\end{equation}
The function $B_1$ behaves as $B_1(4m_q^2/m_a^2)\approx 1$ for $m_q\ll m_a$ and $B_1(4m_q^2/m_a^2)\approx -m_a^2/(12m_q^2)$ for $m_q\gg m_a$. The explicit form of $B_1$ is given in e.g.~\cite{Bauer:2017ris}. For light ALPs, $m_a \ll \Lambda_{\text{QCD}}$, the decay into three pions may be kinematically accessible, with a decay rate which is given in~\cite{Bauer:2017ris,Bauer:2020jbp}.  However, it is worth noting that the FCC-ee program as currently envisioned will not be able to produce significant numbers of ALPs that are heavy enough to decay in two top quarks, due to the high center of mass energy that would be required. Depending on their lifetime, ALPs can decay promptly at the interaction point or they may decay after having travelled a certain distance inside the detector.

\begin{figure}[hbtp]
\centering
\subfigure[]{\includegraphics[width=0.45\textwidth]{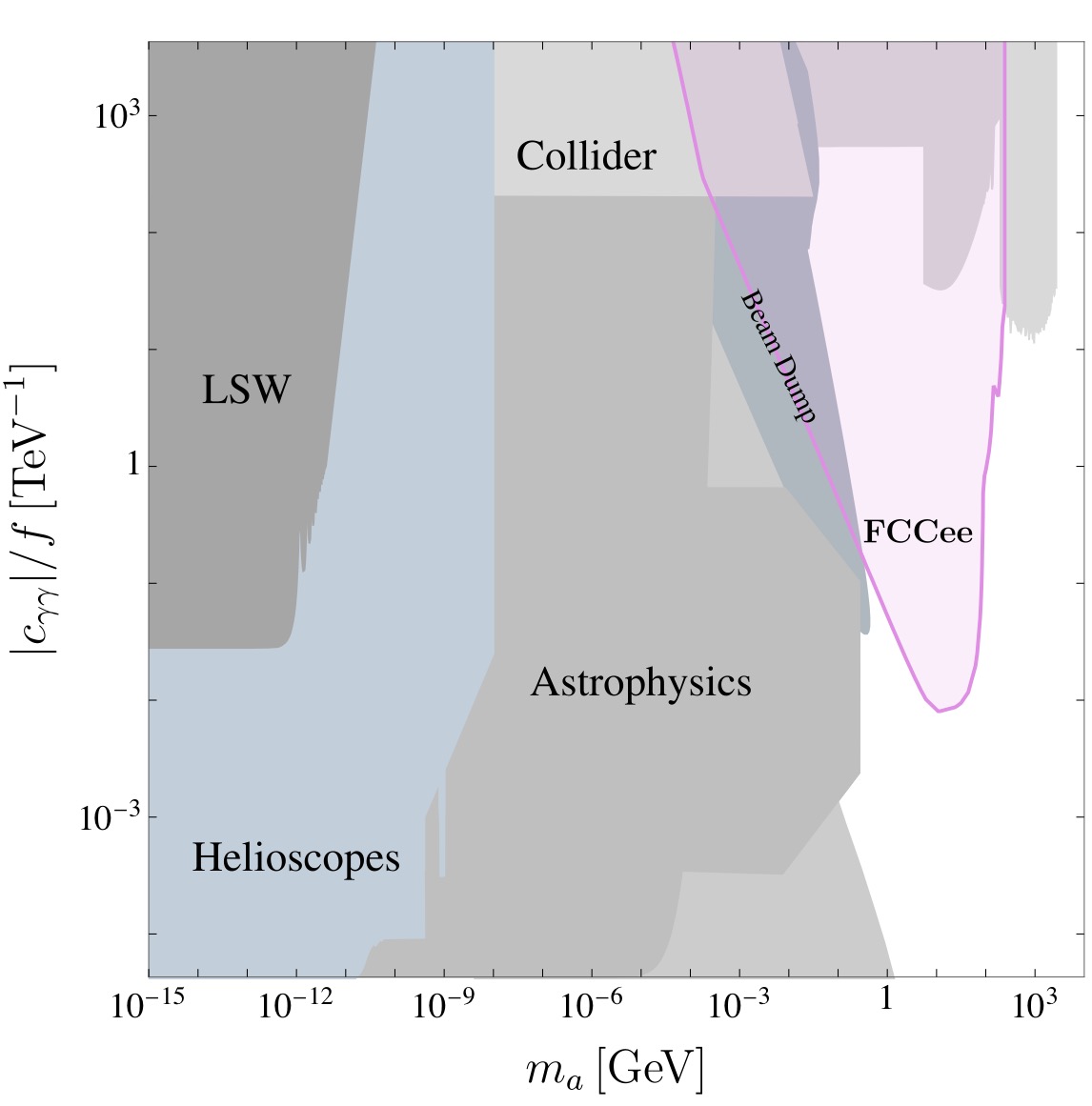}\label{fig:ALPSensitivity_3gamma}} \hspace{1cm}
\subfigure[]{\includegraphics[width=0.45\textwidth]{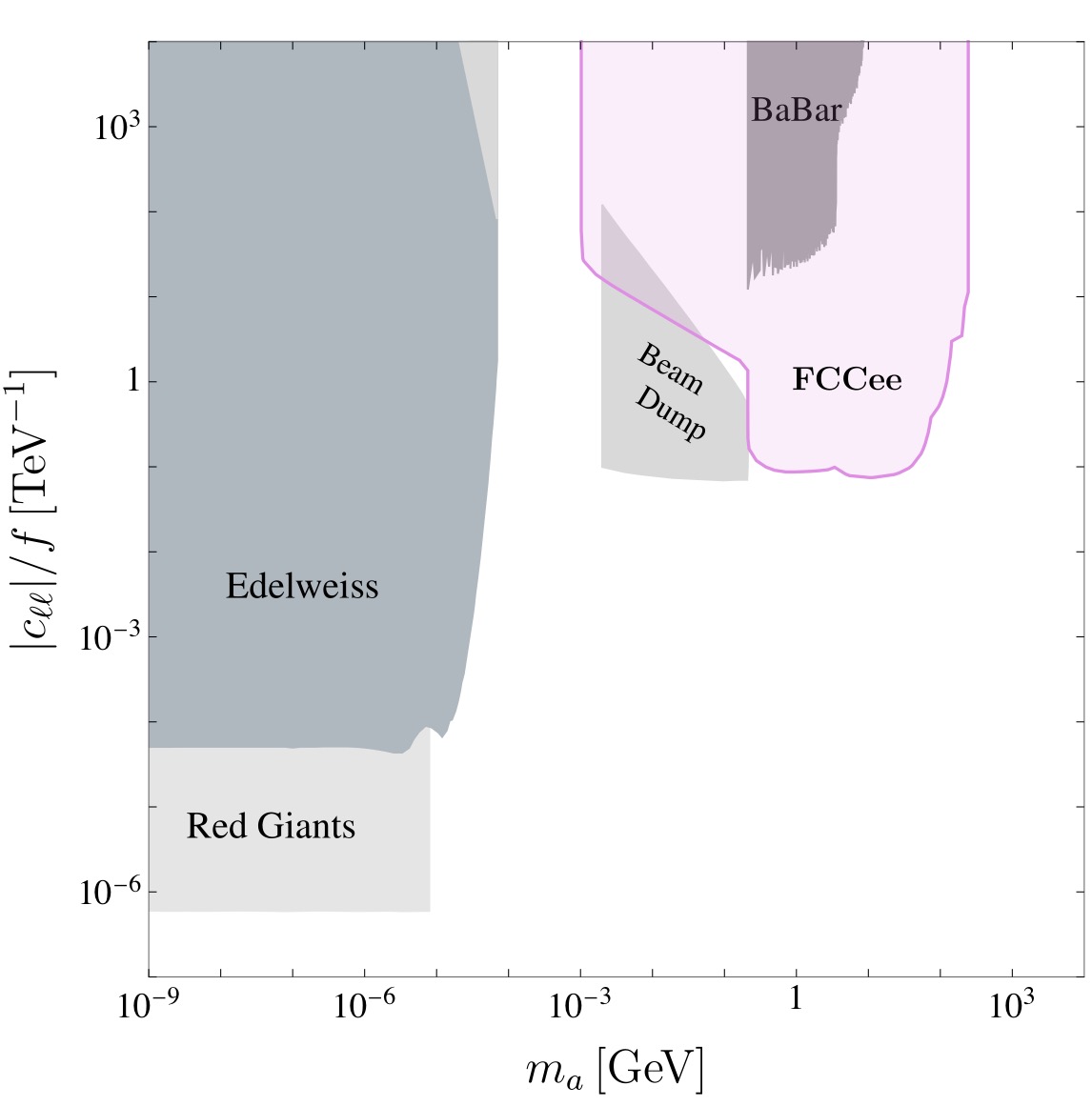}\label{fig:ALPSensitivity_gamma2lep}}
\caption{Projected sensitivity of FCC-ee in (a) $e^+ e^- \to \gamma a \to 3 \gamma$ and (b) $e^+ e^- \to \gamma a \to \gamma \ell^+ \ell^-$ in purple. Figure adapted from Figure~\cite{Bauer:2018uxu}.}
\label{fig:ALPSensitivity}
\end{figure}

At FCC-ee, all combinations of ALP production modes with visible and invisible decay modes can be investigated~\cite{Bauer:2018uxu,Knapen:2021svn}. While many processes, in particular exotic Higgs decays, depend on two independent couplings, under certain assumptions a few processes only depend on a single coupling parameter. For example, the $e^+ e^- \to \gamma a \to 3 \gamma$ and $e^+ e^- \to Z a \to Z \gamma \gamma$ processes only depend on the ALP-photon coupling $c_{\gamma \gamma}$ when it is assumed that both the ALP-photon and the ALP-photon-$Z$ couplings originate from the ALP coupling to either $SU(2)_L$ or $U(1)$ gauge bosons before EW symmetry breaking. If the ALP only couples to $U(1)$ gauge bosons, then $c_{\gamma Z} = -s_w^2 c_{\gamma \gamma}$. In this case, Figure~\ref{fig:ALPSensitivity_3gamma} shows the projected sensitivity of FCC-ee to $c_{\gamma \gamma}$ using the $e^+ e^- \to \gamma a \to 3 \gamma$ channel~\cite{Bauer:2018uxu}. This analysis assumes at least four signal events and combines the $Z$-pole run with runs at $\sqrt{s} = 2m_W$ and $\sqrt{s} = 250\,$ GeV. Further details are provided in~\cite{Bauer:2018uxu}.
Another process that depends only on a single coupling is $e^+ e^- \to \gamma a \to \gamma \ell^+ \ell^-$ when the ALP-photon and the ALP-photon-$Z$ couplings are induced via a loop of leptons. In this case,
\begin{align}
    c_{\gamma \gamma} = \sum_{\ell = e,\mu, \tau} c_{\ell \ell} B_1(4m_\ell^2/m_a^2)  \quad\text{and}\quad c_{\gamma Z} = (s_w^2 - 1/4) c_{\gamma \gamma} \ .
\end{align}
Figure~\ref{fig:ALPSensitivity_gamma2lep} shows the projected sensitivity of FCC-ee to $c_{\ell \ell}$ using the process $e^+ e^- \to \gamma a \to \gamma \ell^+ \ell^-$~\cite{Bauer:2018uxu}. 

The $e^+ e^- \to \gamma a \to 3 \gamma$ and $e^+ e^- \to \gamma a \to \gamma \ell^+ \ell^-$ searches are sensitive to ALP decay lengths of up to $1.5$\,m and $2$\,cm, respectively. The search for long-lived ALPs may be significantly improved with the installation of a dedicated far detector that could probe decay lengths of up to $100\,$m~\cite{Schafer:2022shi,Tian:2022rsi}. For FCC-ee reach on the relaxion, see Ref.~\cite{Frugiuele:2018coc}.
In addition to direct measurements, FCC-ee will be able to significantly constrain the ALP contribution to the oblique parameters~\cite{Bauer:2017ris,Bauer:2018uxu}, whose determination is expected to improve by an order of magnitude~\cite{deBlas:2016ojx}.

\subsection{Exotic Higgs Boson Decays}\label{sec:theory_higgs}
\emph{Editor: Christopher Verhaaren}

The Higgs boson has a unique role within the SM. It is the only apparently elementary scalar particle that has been discovered. In particular, whatever new physics is responsible for the cosmological DM, the apparent asymmetry between matter and antimatter, or small neutrino masses may well have some coupling to the Higgs boson. In short, the Higgs boson is a likely gateway to what lies beyond the SM. 

\begin{figure}[hbtp]
\centering
\includegraphics[width=0.45\textwidth]{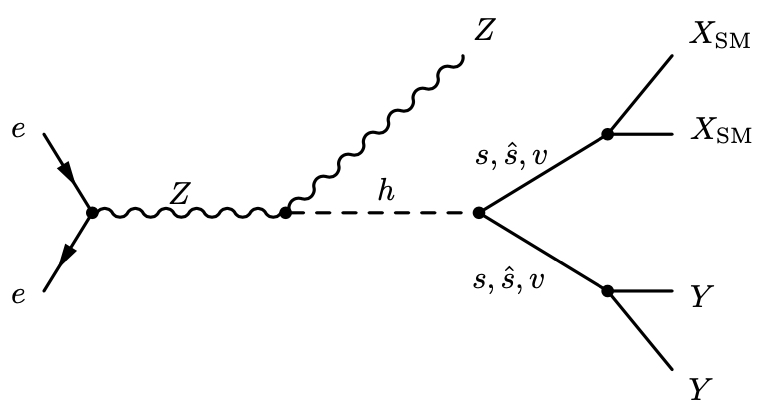}
\caption{Example production of LLPs through exotic decays of the Higgs boson $h$. The Higgs decays to a pair of scalars $s$, pseudoscalars $\hat{s}$, or vectors $v$. At least one of these decays within the detector volume to SM particles. The other may or may not decay within the detector and may decay to visible or invisible states.}
\label{fig:ExoticHiggsFeynmanDiagram}
\end{figure}

The two-body decays of the Higgs boson to SM particles are controlled by small Yukawa couplings or loop suppression making its decay width much smaller than its mass. Consequently, current bounds on the Higgs width leave plenty of room for ``exotic'' decays, that is, decays not predicted by the SM. However, future colliders, like FCC-ee, will be able to measure the Higgs width much more precisely than at the LHC~\cite{deBlas:2019rxi}. The products of these exotic Higgs boson decays can decay promptly themselves or be completely stable, each of which present their own experimental challenges and advantages. However, searches for particles whose lifetimes are more intermediate, i.e., that decay within the experimental detector but at a measurable distance from the interaction point, can have very low backgrounds in comparison to prompt searches. This gives Higgs boson decays to LLPs remarkable power to probe particles and sectors whose couplings to the Higgs are small but nonzero. The $e^+ e^- \to Zh$ processes shown in Figure~\ref{fig:ExoticHiggsFeynmanDiagram} illustrate the utility of the FCC-ee collider. Because the initial state and the $Z$ decays are well understood, invisible, partially invisible, and displaced decays of the Higgs boson can be probed with confidence. For a review, see Ref.~\cite{Curtin:2018mvb} and the recent work in Ref.~\cite{Cepeda:2021rql}.

The characteristics of LLPs vary considerably. Exotic Higgs decays to spin-zero particles are considered first. Such decays at future lepton colliders were considered in Ref.~\cite{Alipour-Fard:2018lsf}. Long-lived scalars may result from simple constructions, such as adding a single scalar field to the SM:
\begin{equation}
    V_\text{scalar}=V_H+V_S+c_1S|H|^2+c_2S^2|H|^2~.
\end{equation}
They may also arise in rich, hidden sectors such as Hidden Valley models~\cite{Strassler:2006im,Strassler:2006ri,Han:2007ae}. Of particular interest are hidden sectors motivated by Neutral Naturalness~\cite{Chacko:2005pe,Burdman:2006tz,Cai:2008au,Craig:2015pha}. These models address the little hierarchy problem through new symmetries, but the symmetry partners of the SM quarks do \emph{not} carry SM color. Instead they are charged under a hidden, QCD-like confining force.

In many models with the long-lived scalar $s$ or pseudoscalar $\hat{s}$, the Higgs boson decay products inherit much of the Higgs' coupling structure. While the actual size of the couplings are reduced by a common small mixing angle $\theta$, the branching fractions are those of a SM Higgs boson with the mass of the LLP. In the scalar case, one often finds
\begin{equation}
    \Gamma\left(s\to X_\text{SM}X_\text{SM}\right)\ =\ \sin^2\theta \ \Gamma\left(h(m_s)\to X_\text{SM}X_\text{SM}\right)~.
\end{equation}
The pseudoscalar case is slightly modified~\cite{Curtin:2013fra,Haisch:2018kqx,Fuchs:2020cmm} and can also include the $h\to \hat{s}\, Z$ decay channel, see for instance the ALP results given in Eqs.~\eqref{eq:widthHZa} and \eqref{eq:widthHaa}. Since the masses of the LLPs must be less than half the Higgs boson mass, the dominant decays modes are into the heaviest kinetically accessible SM quarks. Thus, for Higgs boson decays into spin-zero LLPs, hadronic final states, and $b$-jets especially, are particularly motivated.

Rather than scalars, the LLPs may be spin-half fermions. These can be related to the BAU~\cite{Cui:2014twa} or to Seesaw explanations of the neutrino masses~\cite{Minkowski:1977sc,Glashow:1979nm,Gell-Mann:1979vob,Mohapatra:1979ia,Yanagida:1980xy,Schechter:1980gr,Shrock:1980ct}. The heavy neutrinos $N$ in these models have been shown to have a wide range of possible decay lengths, including within the volume of an FCC-ee detector~\cite{Batell:2016zod,Accomando:2016rpc,Accomando:2017qcs}. The $N$ mainly decay into a SM lepton and an off-shell weak gauge boson. This leads to three-body final states which may be composed of both quarks (jets) and leptons.

The Higgs boson can also decay to long-lived vectors $v$. A simple framework is the Hidden Abelian Higgs model~\cite{Curtin:2014cca}. In this case, a new $U(1)'$ gauge symmetry is broken by a hidden Higgs field $h_D$ that generates a vacuum expectation value (VEV). The hidden photon $A_\mu'$ of the new gauge symmetry gets a mass proportional to the hidden Higgs VEV and can also have kinetic mixing with the SM through
\begin{equation}
    -\frac{\epsilon}{2\cos\theta_W}F_{\mu\nu}'F^{\mu\nu}_Y~,
\end{equation}
where $\theta_W$ is the weak mixing angle and $F^{\mu\nu}_Y$ is the field strength for SM hypercharge. The parameter $\epsilon$ can vary over a huge range, and controls the degree to which SM fermions couple to $A_\mu'$. For sufficiently small $\epsilon$, the massive hidden photon is a LLP. 

The hidden photon's coupling to SM fields is proportional to their hypercharge. This means that, when and if they are kinematically accessible, quark final states make up most of its branching fraction, though decay rates to leptons are non-negligible. A small $\epsilon$ also means that the direct coupling of the hidden photon to the SM Higgs boson is small. However, the mixing between SM Higgs and the hidden Higgs can be larger than $\epsilon$. This allows the Higgs boson to decay to two hidden photons at a larger rate.

In summary, the Higgs boson may have appreciable decay widths into LLPs of various spin. The decay modes of the LLPs can vary, but it has been shown that hadronic final states play a significant role in all the decay types outlined above. Decays to long-lived fermions stand out as different, in that their leading decays are three-body. Pseudoscalars may also lead to $h\to \hat{s}\,Z$ decays, but in general, the $h\to XX$ process captures most of the interesting possibilities. Assuming the $X$ particle has significant branching into SM quarks (and possibly into $b$ quarks in particular) appears to be the most motivated benchmark. Of course, the variety of other decays can be leveraged in more model-specific analyses.

\begin{figure}[t!]
\centering
\includegraphics[width=0.55\textwidth]{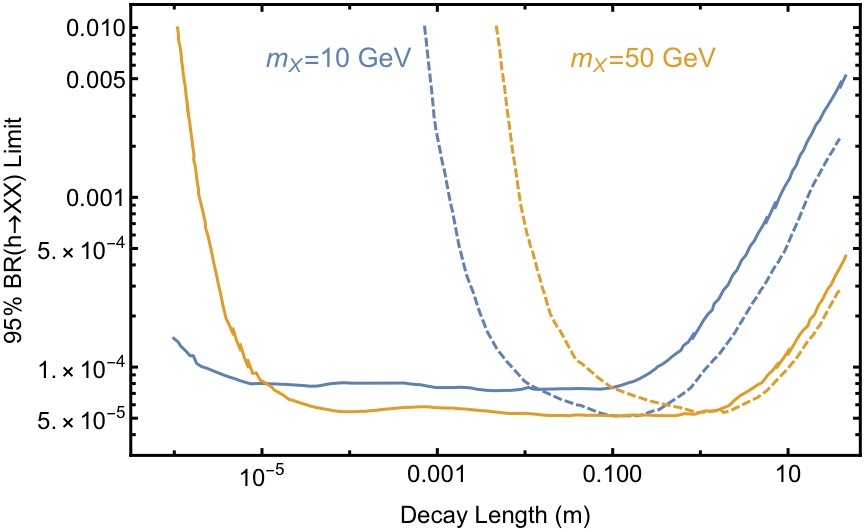}
\caption{Plot of data recorded in~\cite{Alipour-Fard:2018lsf} to  illustrate the potential sensitivity of FCC-ee to exotic Higgs boson decays to LLPs, denoted $X$. Two LLP mass benchmarks are shown: 10 GeV (blue) and 50 GeV (tan). For each benchmark two search strategies are presented. The solid line employs an invariant mass cut to improve sensitivity at shorter decay lengths, the dashed line relies on longer decay lengths to reduce SM backgrounds. }
\label{fig:ExoticHiggsLLPSensitivity}
\end{figure}

Figure~\ref{fig:ExoticHiggsLLPSensitivity} displays an illustration, taken from Ref.~\cite{Alipour-Fard:2018lsf}, of how sensitive FCC-ee can be to Higgs boson decays to long-lived $X$ particles. The 95\% limit on the exotic branching fraction to these particles is plotted as a function of the $X$'s decay length. Two mass benchmarks, $m_X=10$ (blue) and 50 (tan) GeV, are shown (additional benchmarks are considered in Ref.~\cite{Alipour-Fard:2018lsf}) for two search strategies. The solid line corresponds to using an invariant mass cut to retain sensitivity to shorter decay lengths. In contrast, the dashed line depends on longer decay lengths to reduce SM backgrounds.

\FloatBarrier

%% file: experiment.tex
\section{Experimental Outlook} \label{sec:exp}
\emph{Editors: Juliette Alimena, Alain Blondel, Rebeca Gonzalez Suarez, Suchita Kulkarni, Chiara Rizzi, Lovisa Rygaard, Richard Ruiz, Anna Sfyrla, Tanishq Sharma}

This section presents new studies produced for this paper that follow up on the theoretical landscape presented in Section~\ref{sec:theory}.

\subsection{Simulation Details}\label{sec:analysis_setup}

For all signal and background processes, the event generator \texttt{MadGraph5\_aMC@NLO} v3.2.0~\cite{Stelzer:1994ta,Alwall:2014hca} is used to simulate at leading order unpolarized, parton-level $e^+e^-$ collisions at $\sqrt{s}=91$~GeV. For all processes, parton-level events are passed to \texttt{Pythia}~\cite{PYTHIA8} v8.303 to simulate parton showering and hadronization. For each signal benchmark point, $50\times10^{3}$ unscaled events were generated, and for each background process, $10^{7}$--$10^{9}$  unscaled events were generated, depending on the process.

The detector response is simulated with \texttt{Delphes} v3.4.2~\cite{deFavereau:2013fsa}, using the latest Innovative Detector for Electron–positron Accelerators (IDEA) FCC-ee detector concept~\cite{Antonello:2020tzq} card. The IDEA detector comprises of a silicon pixel vertex detector; a large-volume, light short-drift wire chamber surrounded by a layer of silicon micro-strip detectors; a thin, low-mass
superconducting solenoid coil; a pre-shower detector; a dual-readout calorimeter; and muon chambers within the magnet return yoke.

The k4SimDelphes project~\cite{k4simdelphes} converts Delphes objects to the EDM4HEP format~\cite{EDM4hep}, which is the common data format used for the simulation of future colliders. A sophisticated analysis framework has been developed for all FCC analyses using the EDM4hep format. It is based on RDataFrames~\cite{TDataFrame}, where \texttt{C++} code is  compiled in a \texttt{ROOT}~\cite{root} dictionary as ``analysers.'' These are subsequently called in \texttt{Python}. Several external packages, such as \texttt{ACTS}~\cite{acts}, \texttt{FastJet}~\cite{Cacciari:2011ma}, and \texttt{awkward}~\cite{awkward}, are included. 

\medskip
\noindent \emph{Heavy Neutral Leptons}

To study Dirac and Majorana HNLs at FCC-ee, the  processes
\begin{subequations}
\label{eq:sim_HNL}
\begin{align}
\textbf{Majorana}~N &:
    e^+ e^- \to Z \to N \nu_e + N\overline{\nu_e}, \quad\text{with}\quad N \to e^+e^-\nu_e + e^+e^-\overline{\nu_e},
    \label{eq:sim_Majorana}
    \\
\textbf{Dirac}~N &:
    e^+ e^- \to Z \to N \overline{\nu_e} + \overline{N}\nu_e, \quad\text{with}\quad N~(\overline{N}) \to e^+e^-\nu_e~(\overline{\nu_e}),
    \label{eq:sim_Dirac}    
\end{align}
\end{subequations}
are simulated using the \texttt{HeavyN}~\cite{Alva:2014gxa,Degrande:2016aje} and \texttt{HeavyN\_Dirac}~\cite{Degrande:2016aje,Pascoli:2018heg} 
Universal \texttt{FeynRules} Object~\cite{Christensen:2008py,Degrande:2011ua,Alloul:2013bka} libraries in conjunction with \texttt{MadGraph5\_aMC@NLO}.
These libraries implement the interaction Lagrangian described in Section~\ref{sec:theory_nu} for Majorana and Dirac $N$, respectively.
A representative subset of Feynman diagrams common to both the Dirac and Majorana case is shown in Figure~\ref{fig:diagram_eeZ_heavyN}. 
For the Majorana case, both LNC and LNV channels are included.
The Dirac case only permits LNC channels. The preservation of spin correlation in the production and decay of $N$ with this setup was checked in Ref.~\cite{Ruiz:2020cjx}. When unspecified, the results consider the Majorana case. As a further benchmark, the assumption that $N$ couples only to the electron-flavor sector is made, i.e.,  $\vert V_{eN}\vert $ is kept nonzero and set $\vert V_{\mu N}\vert, \vert V_{\tau N}\vert = 0$. Only one heavy neutrino mass eigenstate is considered. SM inputs are fixed according to the values in Ref.~\cite{Degrande:2016aje}.

\medskip
\noindent \emph{Axion-like particles}

To study the production ALPs $a$ from $Z$ decays at FCC-ee, the process
\begin{equation}
    \textbf{ALP} : e^+e^- \to Z \to a\gamma, \quad\text{with}\quad a \to \gamma\gamma,
\end{equation}
is simulated using the model libraries of Ref.~\cite{ALPs} in conjunction with \texttt{MadGraph5\_aMC@NLO}. These libraries implement the Lagrangian described in Section~\ref{sec:theory_alps}.

\medskip
\noindent \emph{Exotic Higgs boson decays}

A simulation study of exotic Higgs decays into LLPs is left for a future paper, as well as additional detector concepts, namely the CLIC-like detector (CLD) design~\cite{Bacchetta:2019fmz}.

\FloatBarrier

\subsection{Heavy Neutral Leptons}\label{sec:analysis_hnl}

Although the most promising Seesaw models feature two or three HNL states in the same mass range, and possibly almost degenerate, a reasonable experimental approach is to begin by considering the production and decay of a single HNL particle.

The branching fraction of a $Z$ boson decay into any light neutrino or antineutrino and a heavy neutrino $N$,  which mixes with the three families of neutrinos is given by~\cite{Ruiz:2015gsa,Pascoli:2018heg}:
\begin{equation}
    {\rm BR}(Z\rightarrow \nu N) = \frac{2}{3}|U_N|^2 \  
    {\rm BR}(Z\rightarrow {\rm invisible})\  
   \left(1+\frac{{m_N}^2}{{2m_Z}^2}\right)
     \left(1-\frac{{m_N}^2}{{m_Z}^2}    \right)    , 
     \label{eq:ZBRvN} 
\end{equation}
where $|U_N|^2 \equiv \sum_{\ell = e, \mu, \tau} |U_{\ell N}|^2 $
is the sum of the mixing matrix elements of the HNL $N$ with the three active neutrinos $\nu_\ell$. As the HNL masses considered here are much heavier than the tau lepton, the total charged current decay rate of the HNL $N\rightarrow \ell_\lambda W^* $  is also  proportional to the same combination of mixing angles. 
\begin{equation}
    \Gamma_N = \frac{1}{c\tau_N}\simeq C_0 C_{MD} |U_N|^2 \  \left(\frac{m_N}{\rm 50 GeV}\right)^5\  \times \  \left(\frac{3.10^9}{\rm 1~cm}   \right)
    \label{eq:Ndecayrate}
\end{equation}
Here, $C_0$ is a numerical coefficient of ${\cal{O}} (1) $ that takes into account the open charged- and neutral-current decays of the heavy neutrino, and   $C_{MD}$ is a coefficient that depends on the Dirac ($C_{MD}=1$) or Majorana ($C_{MD}=2$) nature of the particle, since twice as many decay channels are open for the Majorana particle decay. Potentially, with sufficient statistics, the direct comparison of the event rate with the lifetime for an HNL of a known mass would allow a discrimination between a Dirac and a Majorana particle.   

\begin{figure}[t!]
\centering
\includegraphics[width=0.8\textwidth]{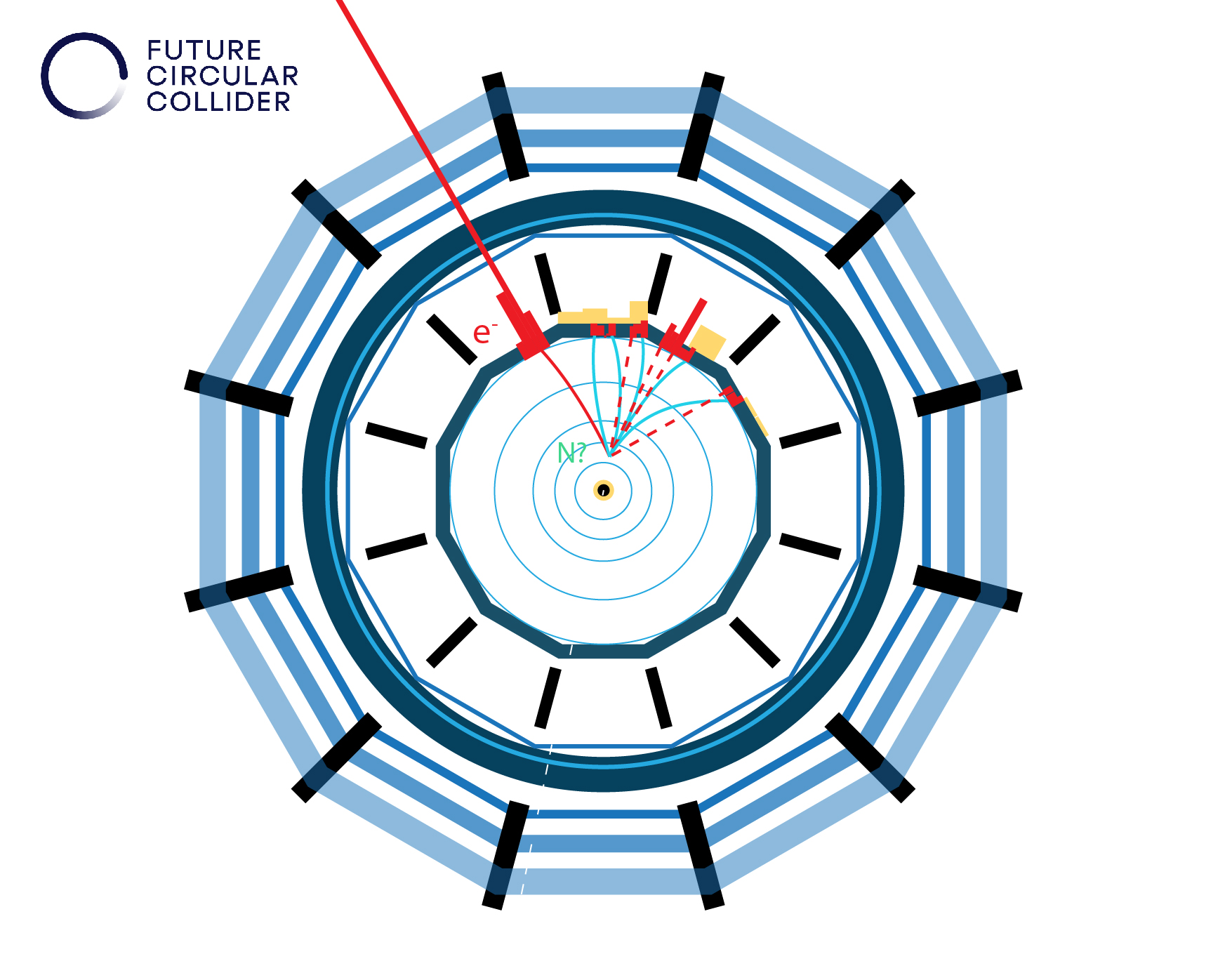}
\caption{Representation of an event display at an FCC-ee detector of a HNL decay into an electron and a virtual $W$ decaying hadronically. Courtesy of the FCC collaboration.}
\label{fig:display}
\end{figure}

The corresponding decay length is then of the order of a meter for a 50 GeV HNL. In those  conditions, a HNL would decay in the volume of an FCC-ee detector, leading to the observable signature of a displaced vertex, with a significant time delay (several nanoseconds) with respect to ultra relativistic particles. This leads to a particularly clean signature, for which a first analysis~\cite{Blondel:2014bra} argued that it could be a background-free search, at least for the dominant charged current decay $N\rightarrow \ell W^* \rightarrow \ell {\rm + hadrons}$. Figure~\ref{fig:display} shows what such a possible decay of the $N$ at a future FCC-ee experiment would look like, in this case for a semileptonic final state.

Furthermore, for $Z \to N \nu_\ell$ decays, the two-body $Z$ decay kinematics results in a mono-chromatic HNL.

Therefore, even in cases where a full, final-state reconstruction is not possible, a simultaneous measurement of  the decay path and of the time-of-flight provides a determination of both the mass and proper decay time on an event-by-event basis. 
A detailed simulation of the process is thus of great interest to understanding how much statistics are required, first to establish the existence of the new particle, and then to establish the possible existence of a lepton number violating process (Majorana vs Dirac nature). This also leads to the identification of specific detector requirements to optimize the discovery potential. 

\FloatBarrier
\subsubsection{Production and Kinematics of Electroweak-scale HNLs}

\begin{figure}[ht!]
\centering
\subfigure[]{\includegraphics[width=0.48\textwidth]{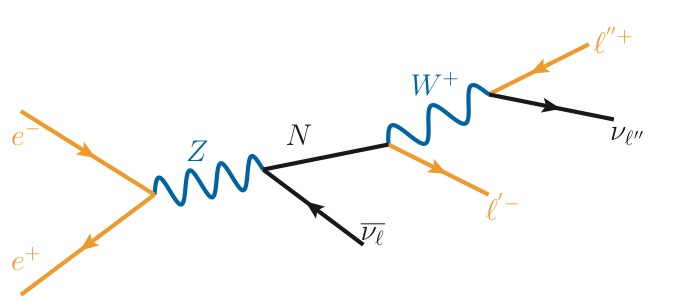}}
\subfigure[]{\includegraphics[width=0.45\textwidth]{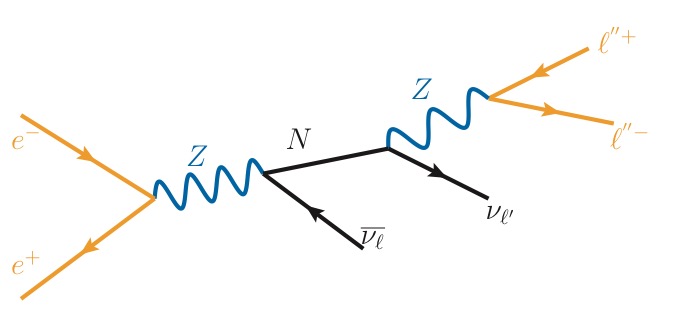}}
\caption{Representative diagrams depicting the $e^+e^- \to Z \to N \nu_\ell$ process at leading order, with $N$ decaying via (a) charged current and (b) neutral current channels to the two-neutrino, two-charged lepton final state.}
\label{fig:diagram_eeZ_heavyN}
\end{figure}

As a first step to exploring the sensitivity of FCC-ee to EW-scale HNLs, Table~\ref{tab:HNLexpectedEvents} shows the cross section (center column) and the expected number of events (right column) for an HNL with a mass of $m_N=50$~GeV when produced and decayed through the process described in Eq.~\eqref{eq:sim_HNL} and shown in Figure~\ref{fig:diagram_eeZ_heavyN}. 

Results are shown for several choices of active-sterile mixing $\vert V_{eN}\vert$, and assume that an integrated luminosity of 150 ab$^{-1}$ is collected during the Tera-$Z$ run of FCC-ee~\cite{FCC-CDR2}. 
No event selection is applied at this stage.

\begin{table}[!ht]
\caption{The cross section and expected number of events at 150 ab$^{-1}$, for an HNL with a mass of $50$~GeV and for several choices of $\vert V_{e N}\vert$. No event selection is applied.}
\label{tab:HNLexpectedEvents}
\centering
\begin{tabular}{|c|c|c|}
\hline
Active-sterile 
& Cross Section  & Expected events  \\
mixing~$\vert V_{e N}\vert$
& [pb] & at 150 ab$^{-1}$ \\ \hline
$1\times 10^{-1}$                  & 2.29                       & 343,200,000            \\ \hline
$1\times 10^{-2}$                  & $2.29\times 10^{-2}$                    & 3,432,000              \\ \hline
$1\times 10^{-3}$                  & $2.29\times 10^{-4}$                    & 34,320                 \\ \hline
$1\times 10^{-4}$                  & $2.29\times 10^{-6}$                    & 343                    \\ \hline
$1\times 10^{-5}$                  & $2.29\times 10^{-8}$                    & 3                      \\ \hline
$1\times 10^{-6}$                  & $2.29\times 10^{-10}$                   & 0                      \\ \hline
\end{tabular}
\end{table}
\FloatBarrier

\begin{figure}[!t]
\centering
\subfigure[]{\includegraphics[width=0.45\textwidth]{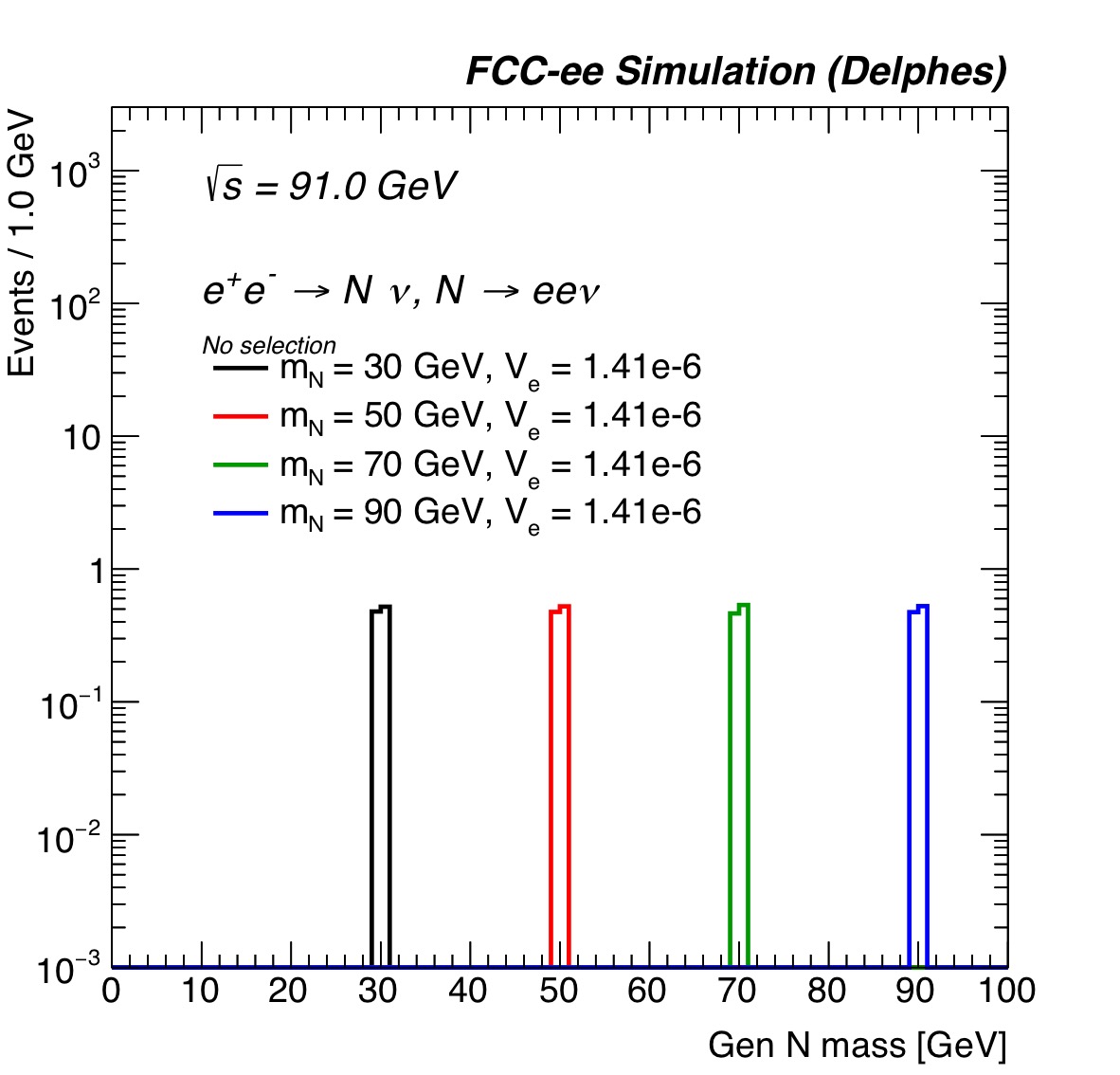}\label{fig:HNLkinematics_mass}}
\subfigure[]{\includegraphics[width=0.45\textwidth]{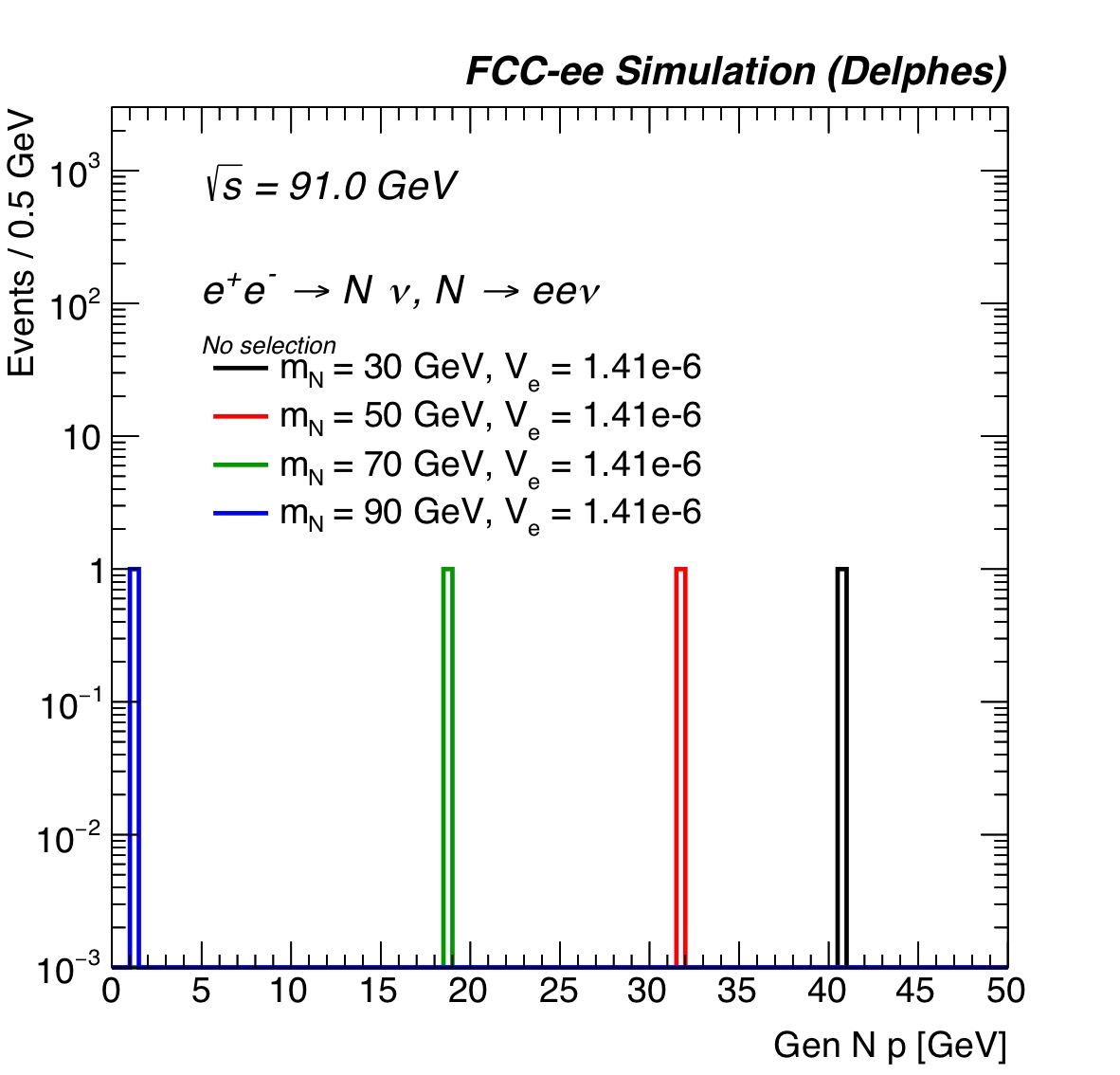}\label{fig:HNLkinematics_mom}}
\subfigure[]{\includegraphics[width=0.45\textwidth]{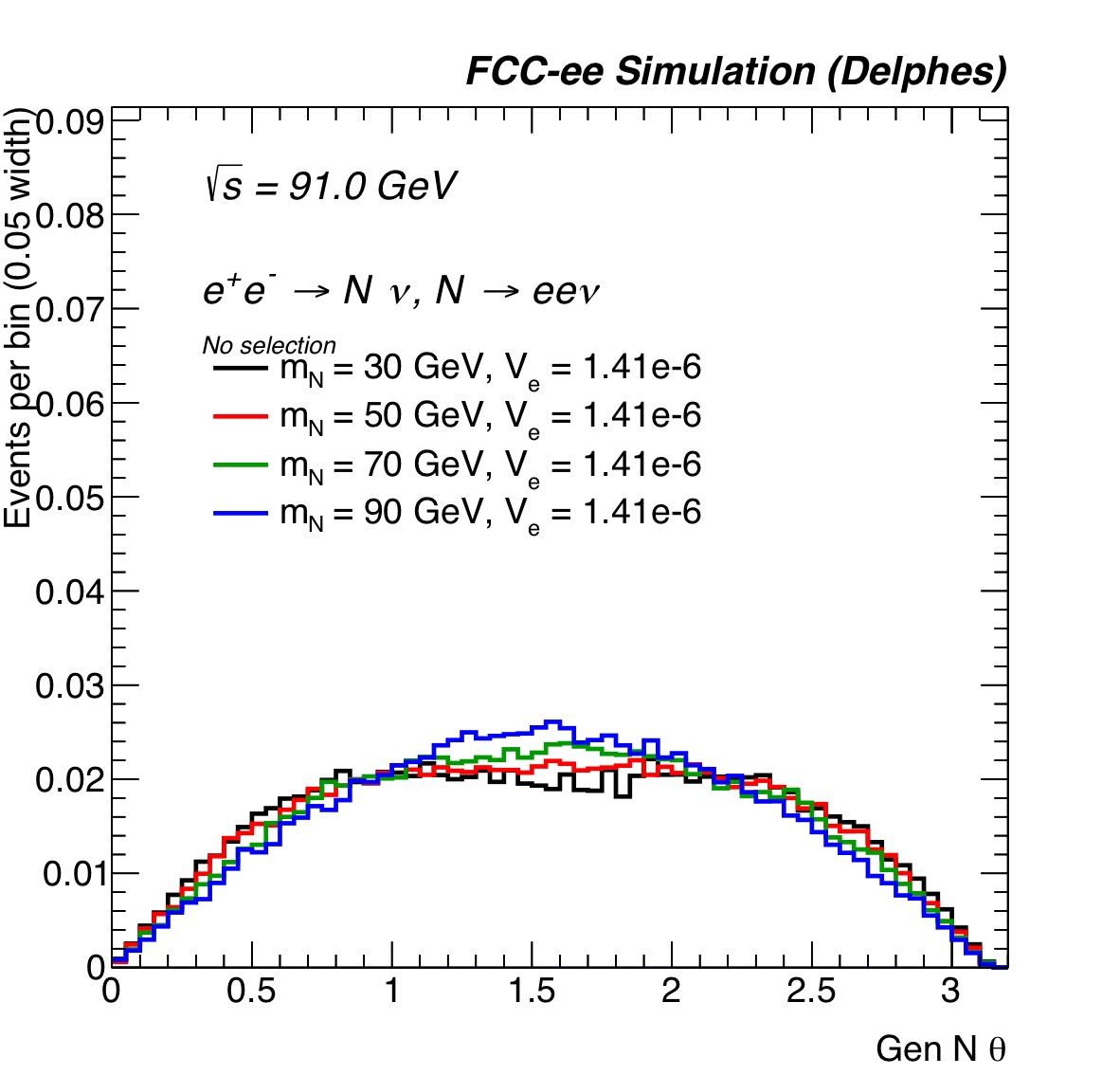}\label{fig:HNLkinematics_polar}}
\caption{For the processes $e^+e^- \to N \nu_e+N \overline{\nu_e}$ with $N\to e^+e^-\nu_e + e^+e^-\overline{\nu_e}$ at $\sqrt{s}=91$~GeV, the generator-level distributions of (a) the invariant mass of $N$, (b) the magnitude of $N$'s three-momentum in the lab frame, and (c) the polar angle of $N$ with respect to the beam axis in the lab frame are shown,
for representative HNL masses and representative active-sterile mixing $\vert V_{eN}\vert = 1.41 \times 10^{-6}$. The distributions are normalized to unit area.
}
\label{fig:HNLkinematics}
\end{figure}

The kinematics of HNLs in the $m_N =20-90$~GeV mass range at FCC-ee can also be studied. Figure~\ref{fig:HNLkinematics} shows the baseline kinematics distributions of $N$ when 
no event selection is applied at this stage.
Here and below, active-sterile mixing of $\vert V_{eN}\vert = 1.41 \times 10^{-6}$ for representative masses of $m_N = 30~(50)~[70]~\{90\}$~GeV is assumed.

Figure~\ref{fig:HNLkinematics_mass} shows the generator-level invariant mass of the HNL, which aligns with the pole mass of $N$. In Figure~\ref{fig:HNLkinematics_mom}, the magnitude of the normalized, generator-level three-momentum $\vert \vec{p}_N\vert$ in the lab frame is presented. From elementary kinematics, $\vert \vec{p}_N\vert$ is given analytically for a massless electron by the formula
\begin{align}
    \vert \vec{p}_N\vert  = \frac{M_Z}{2}\left(1-\frac{m_N^2}{M_Z^2}\right).
    \label{eq:momentumN}
\end{align}
This corresponds to $\vert \vec{p}_N\vert\approx 40.7~(31.9)~[18.7]~\{1.2\}$~GeV for the representative $m_N$ under consideration and is in good agreement with the values $\vert \vec{p}_N\vert\approx 41~(32)~[19]~\{1.2\}$~GeV shown in Figure~\ref{fig:HNLkinematics_mom}.
Finally, the generator-level polar angle $\theta$ of $N$ with respect to the beam axis in the lab frame is presented in Figure~\ref{fig:HNLkinematics_polar}. The distribution shows that a bulk of events feature central ($0.5 < \theta < 2.5$) HNLs, as one would expect from a high-$p_T$ process.

\begin{figure}[!t]
\centering
\subfigure[]{\includegraphics[width=0.45\textwidth]{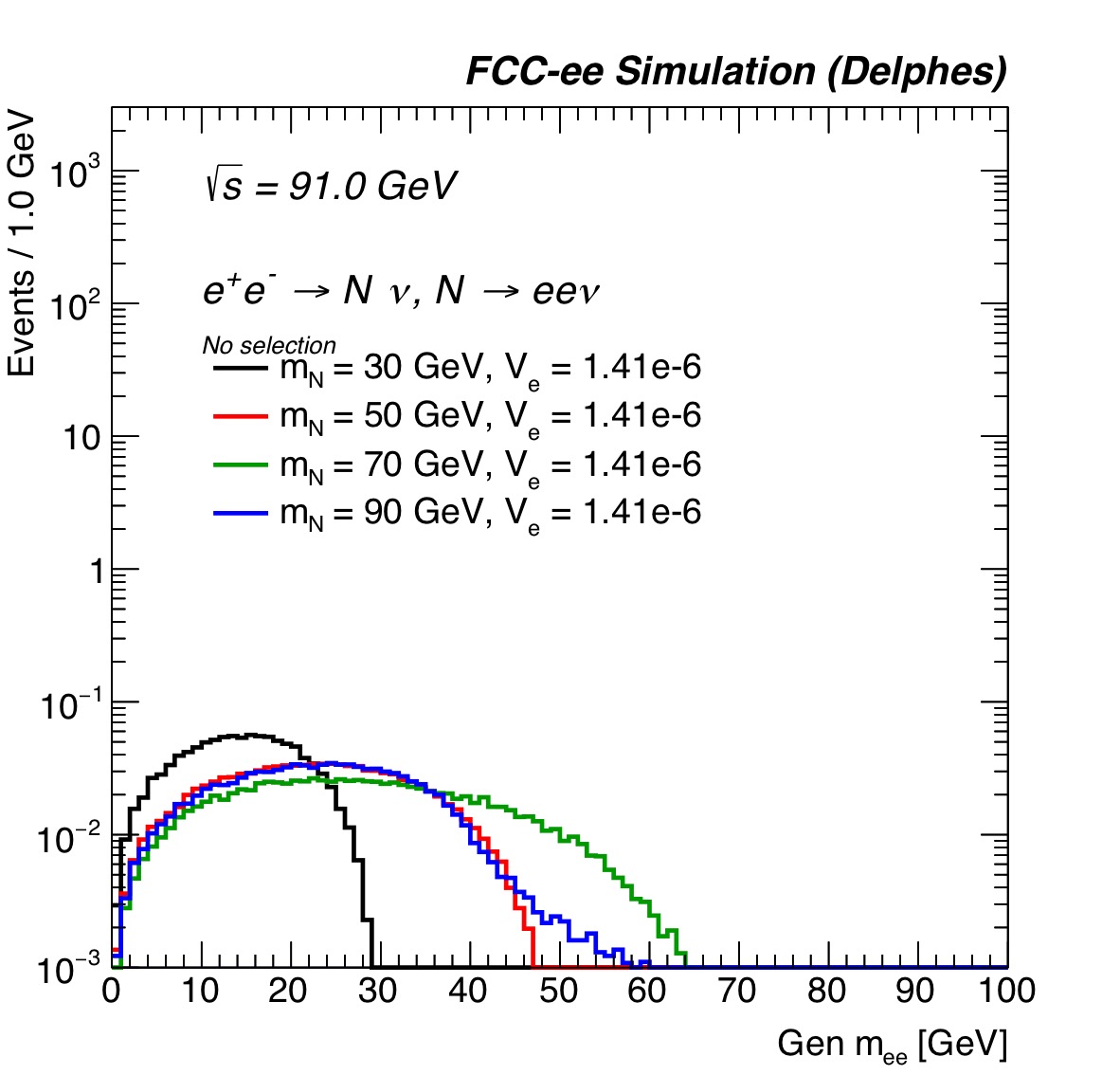}\label{fig:HNLinvMass_gen}}
\subfigure[]{\includegraphics[width=0.45\textwidth]{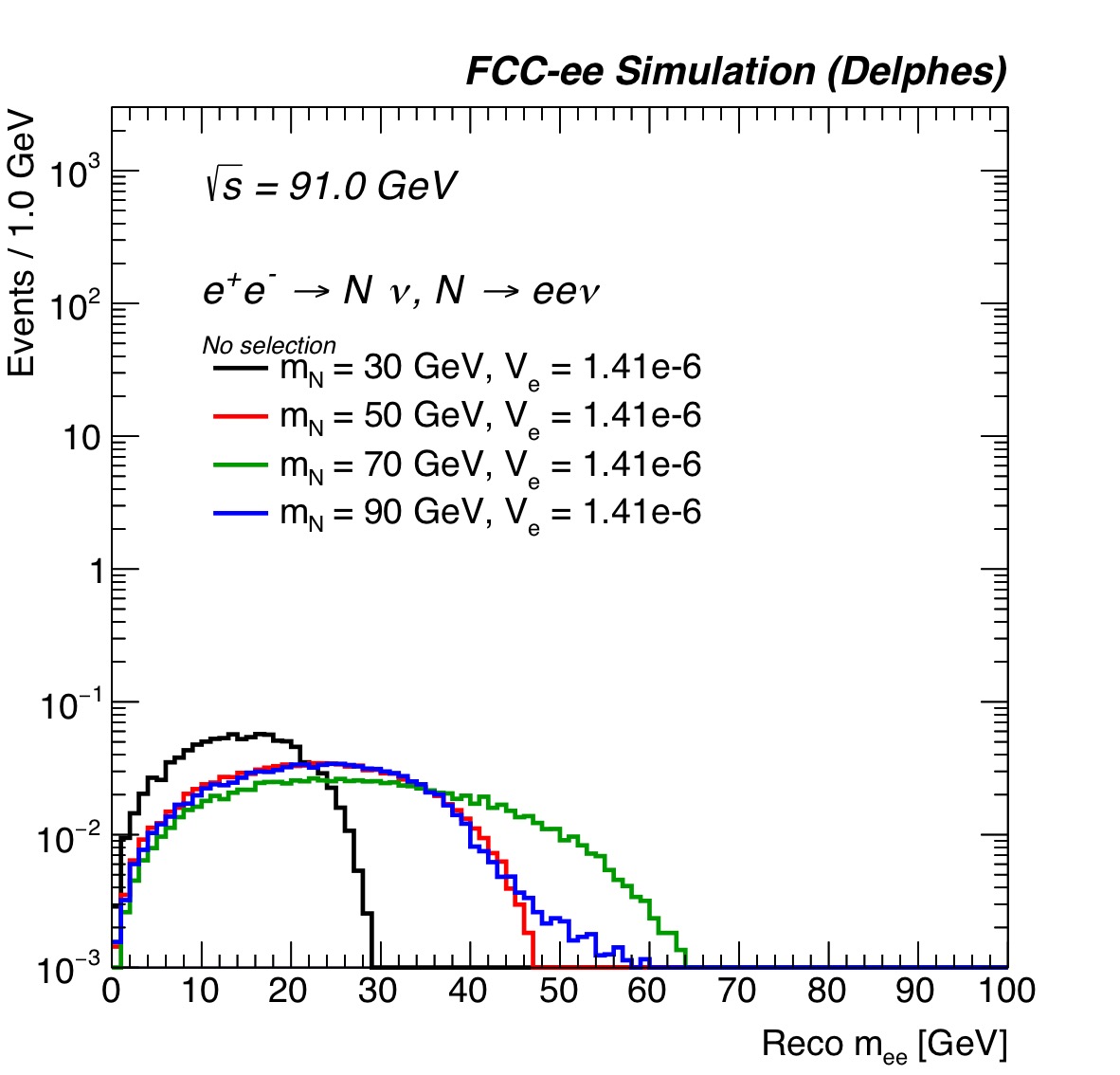}\label{fig:HNLinvMass_reco}}
\caption{For the same processes and benchmark mass and $\vert V_{eN}\vert$ choices as in Figure~\ref{fig:HNLkinematics}, the differential distributions with respect to the invariant mass of the $(e^+e^-)$ system
$m_{ee}$ at (a) the generator level and (b) after reconstruction. No selection criteria have been applied. The distributions are normalized to unit area.
}
\label{fig:HNLinvMass}
\end{figure}

To explore the potential impact of finite detector resolution, limited geometric coverage, and detector mismeasurements, Figure~\ref{fig:HNLinvMass} shows the  distributions with respect to the invariant mass of the $(e^+e^-)$ system, which is given for massless electrons by the formula 
\begin{align}
    m_{ee} = \sqrt{(p_{e^+}+p_{e^-})^2} \approx \sqrt{2p_{e^+}\cdot p_{e^-}}
    = \sqrt{2E_{e^+}E_{e^-}(1-\cos\theta_{ee})}
    ,\
    \label{eq:hnl_mee_def}
\end{align}
at (a) the generator level (Gen) and (b) the reconstruction level (Reco).
In both cases, no selection criteria have been applied and the same representative inputs as above are assumed.

Consider first the generator-level case in Figure~\ref{fig:HNLinvMass_gen}.
As both charged leptons in the final state originate from the $N\to e^+e^+X$ decay, the distribution of $m_{ee}$ is dictated by the properties of $N$ itself. For instance: for each of the mass benchmarks, the value of the observable $m_{ee}$ does not exceed $m_N$ itself, i.e., $\max(m_{ee}) < m_N$. This can be understood from momentum conservation:
\begin{align}
    m_N^2 = (p_{e^+}+p_{e^-}+p_\nu)^2 = p_\nu^2 + 2(p_\nu\cdot p_{e^+})+ 2(p_\nu\cdot p_{e^-}) + m_{ee}^2 \gtrsim m_{ee}^2\ 
    .
\end{align}
When $m_{ee}$ is close to $m_N$, one can infer that the final-state neutrino carries little-to-no energy. For $m_N=90$~GeV, kinematic peculiarities arise due to threshold effects. More specifically, since $\sqrt{s}=91$~GeV, one can consider $N$ to be essentially at rest when $m_N=90$~GeV. For such masses, the two-body decay $N\to e^\pm W^\mp$ becomes kinematically favored. The energy of this first electron and $W$ are given approximately by formulae similar to Eq.~\eqref{eq:momentumN}, and come out to be 
$E_1\approx9.4$~GeV and $E_W\approx80.6$~GeV.

Assuming that the decay products of the $W$ boson are configured in the lab frame such that the second electron carries away all the energy of $W$, i.e., $E_2\approx E_W$, then the formula for $m_{ee}$ shows that the maximum invariant mass for $m_N=90$~GeV is about $\max(m_{ee})\approx \sqrt{4(9.4~\mathrm{GeV})(80.6~\mathrm{GeV})}\approx55$~GeV. This is in agreement with Figure~\ref{fig:HNLinvMass_gen}.

Comparing Figs.~\ref{fig:HNLinvMass_gen} and \ref{fig:HNLinvMass_reco} demonstrates some impact of the event reconstruction. Importantly, many of the kinematic features found at the generator level survive at the reconstruction level. In particular, the endpoints of $m_{ee}$ are preserved. Likewise, the means of each distribution, which span about $m_{ee}^{\rm mean}\approx 14-28$~GeV, remain unaltered at the reconstruction level. The relatively small impact of reconstruction effects can be tied to the high requirements of the FCC sub-detector systems.

\begin{figure}[!t]
\centering
\subfigure[]{\includegraphics[width=0.45\textwidth]{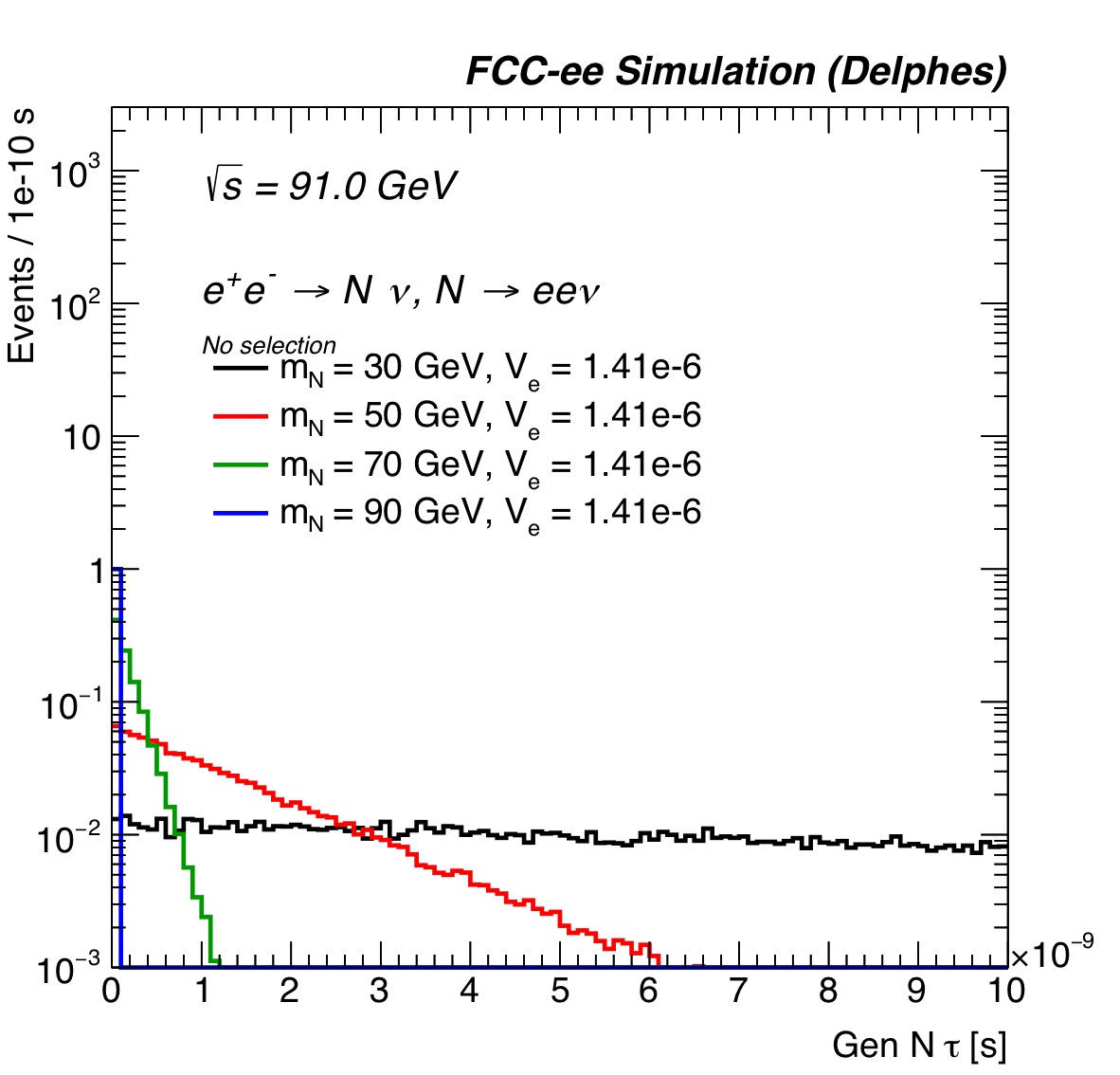} \label{fig:HNLlifetime_gen}}
\subfigure[]{\includegraphics[width=0.45\textwidth]{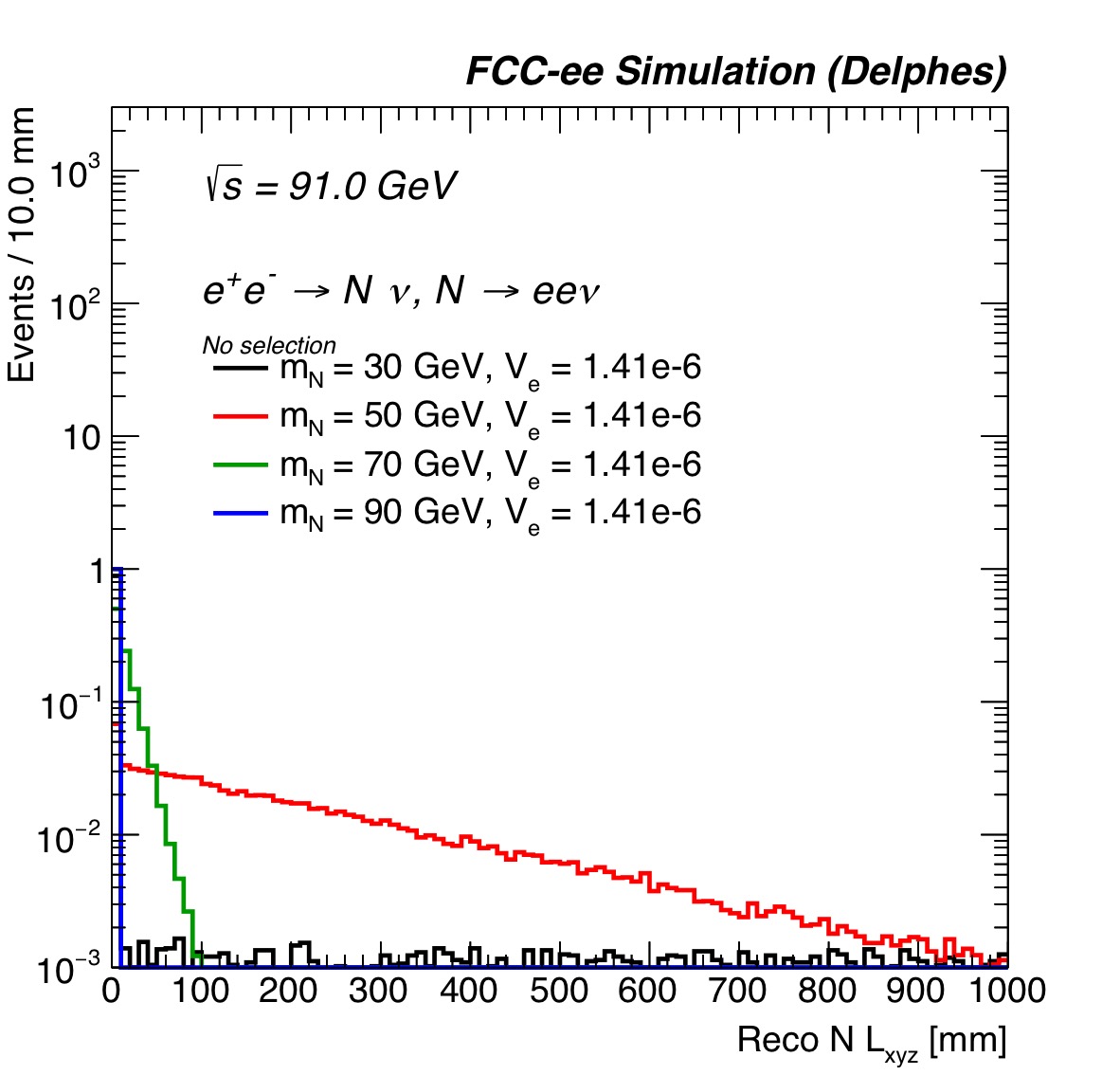} \label{fig:HNLlifetime_reco}}
\subfigure[]{\includegraphics[width=0.45\textwidth]{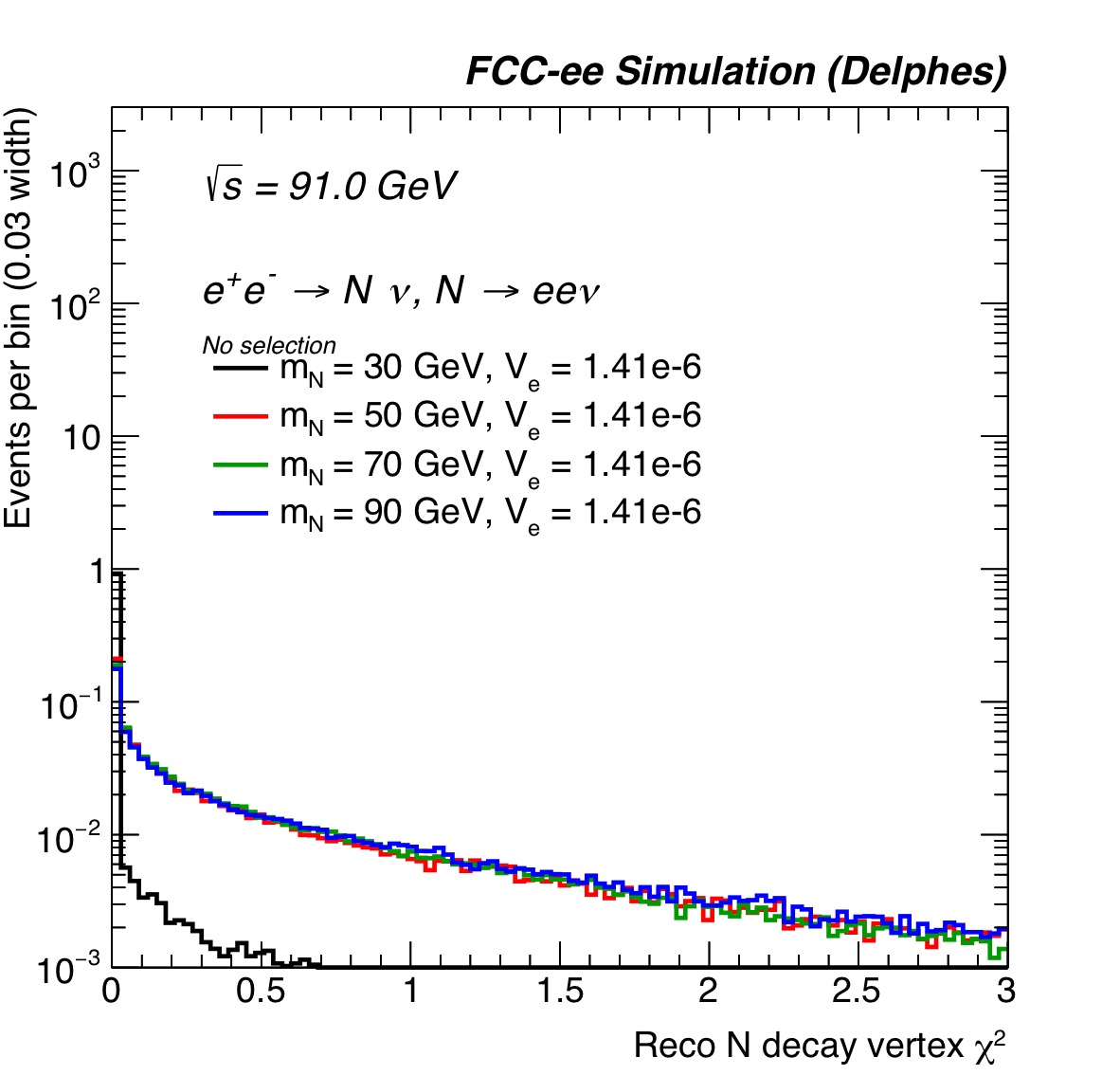}
\label{fig:HNLlifetime_chi2}}
\caption{For the same processes and benchmark mass and $\vert V_{eN}\vert$ choices as in Figure~\ref{fig:HNLkinematics}, the differential distributions with respect to 
(a) the generator-level lifetime of $N$ in the lab frame;
(b) the reconstruction-level three-dimensional decay length of the $N$;
and
(c) the $\chi^2$ of the reconstructed decay vertex of the HNL are shown. 
No selection criteria have been applied. The distributions are normalized to unit area.
}
\label{fig:HNLlifetime}
\end{figure}

In the absence of additional new physics, HNLs with masses below the EW scale and active-sterile mixing much smaller than unity are generically long-lived. To explore this at FCC-ee, Figure~\ref{fig:HNLlifetime} shows (a) the generator-level lifetime (s) of $N$, given by $\tau = \gamma_N \tau_N$, where $\gamma_N = E_N/m_N$ is the Lorentz boost of $N$ in the lab frame, and $\tau_N$ is the proper lifetime; (b) the reconstructed three-dimensional decay length (mm) of the HNL $(L_{xyz})$;
and
(c) the $\chi^2$ of the reconstructed displaced vertex.

For a fixed width of $\vert V_{eN}\vert = 1.41\times10^{-6}$, different qualitative features can be observed for the representative $m_N$. For instance, at $m_N=30$~GeV, characteristic generator-level lifetimes readily exceed several seconds. This implies displaced vertices can be well beyond one or more meters, and therefore outside the fiducial coverage of the IDEA detector. In these instances, a large region of the event's phase space corresponds to long-lived HNLs that ostensibly appear as missing momentum.

For heavier $N$, lifetimes are drastically smaller, with most HNL events exhibiting a lifetime of less than $1-2~s$  for $m_N\gtrsim 50$~GeV. For $m_N=50~(70)$~GeV, such lifetimes correspond to reconstructed displacements that are mostly within $L_{xyz}=50~(100)$~mm. Finally, in Figure~\ref{fig:HNLlifetime_chi2}, the $\chi^2$ curves indicate that the displaced vertices are well-reconstructed, with small $\chi^2$ values.

\FloatBarrier

\subsubsection{Backgrounds and Event Selection}

Several backgrounds to the HNL processes described in Eq.~\ref{eq:sim_HNL} are considered, namely, $Z$ bosons that decay to electron-positron pairs, to tau pairs, to light quarks, to charm quark pairs, and to $b$ quark pairs. These background processes were simulated with the conditions described above. 

\begin{figure}[!t]
\centering
\includegraphics[width=0.45\textwidth]{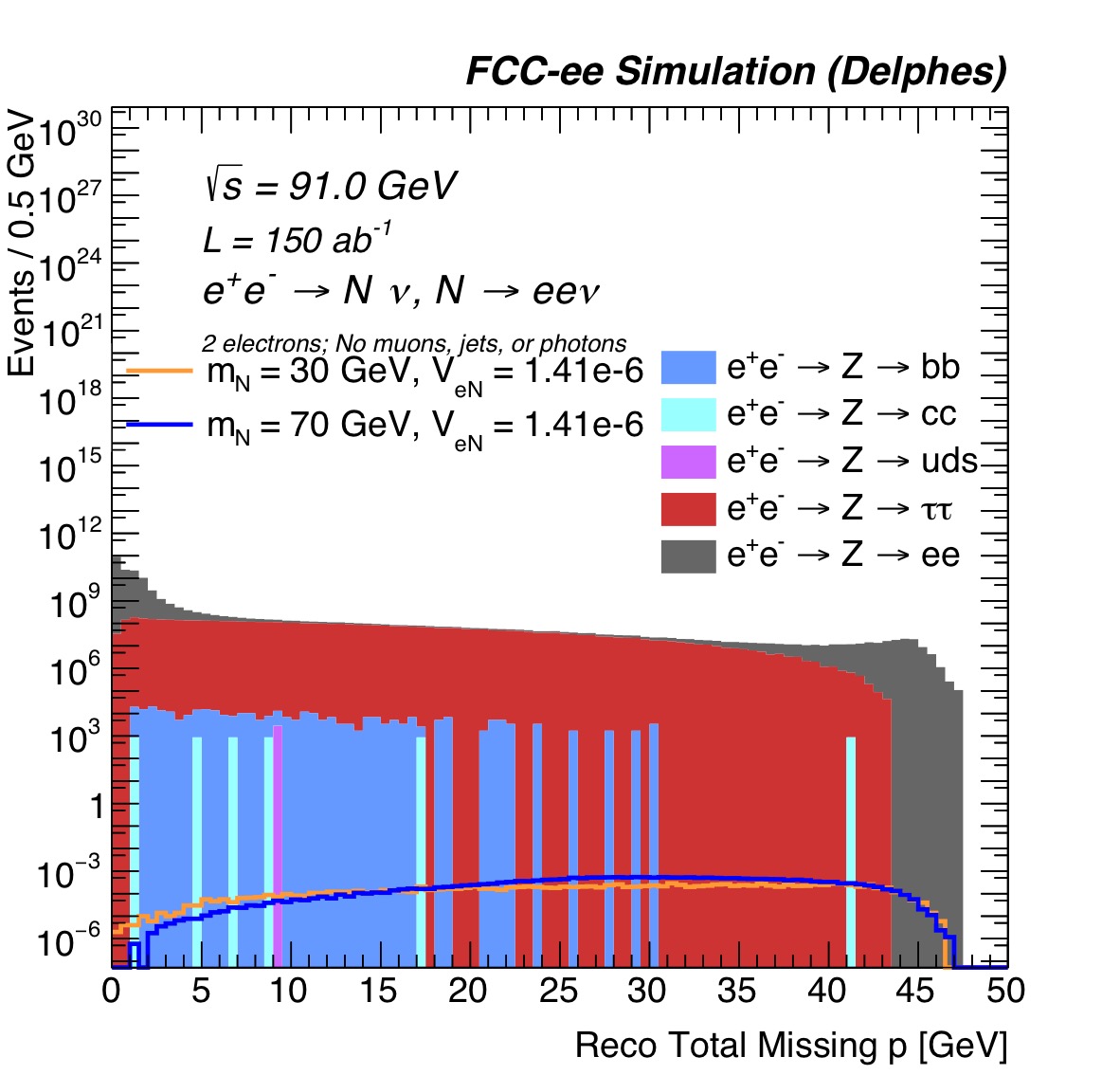}
\caption{The normalized, reconstructed-level total missing momentum, for representative HNL signal benchmark mass and $\vert V_{eN}\vert$ choices, as well as background processes. Exactly two reconstructed electrons are required, as well as that there are no reconstructed muons, jets or photons in each event.
}
\label{fig:HNLmissingEnergy}
\end{figure}

\begin{figure}[!t]
\centering
\subfigure[]{\includegraphics[width=0.45\textwidth]{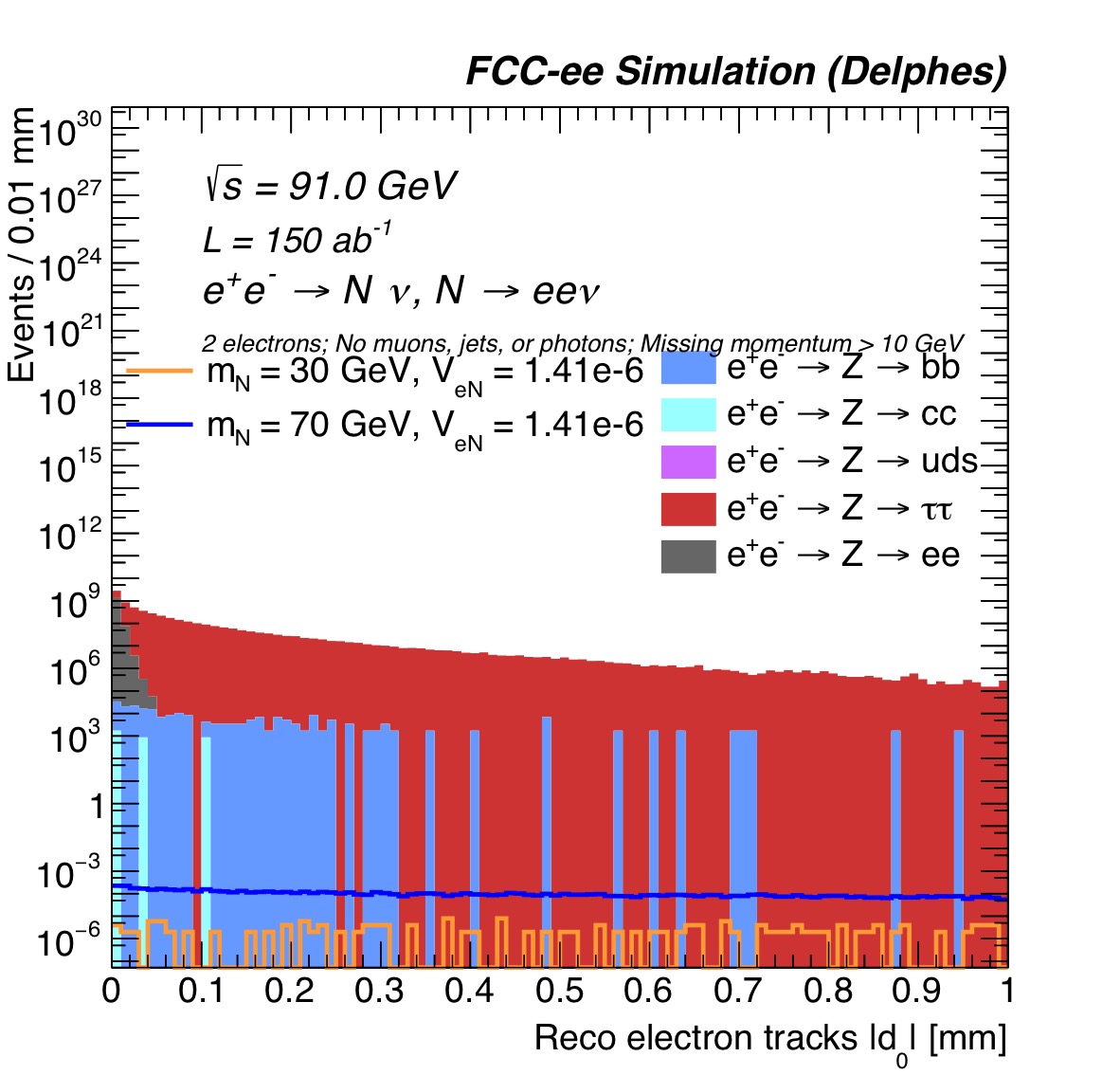}}
\subfigure[]{\includegraphics[width=0.45\textwidth]{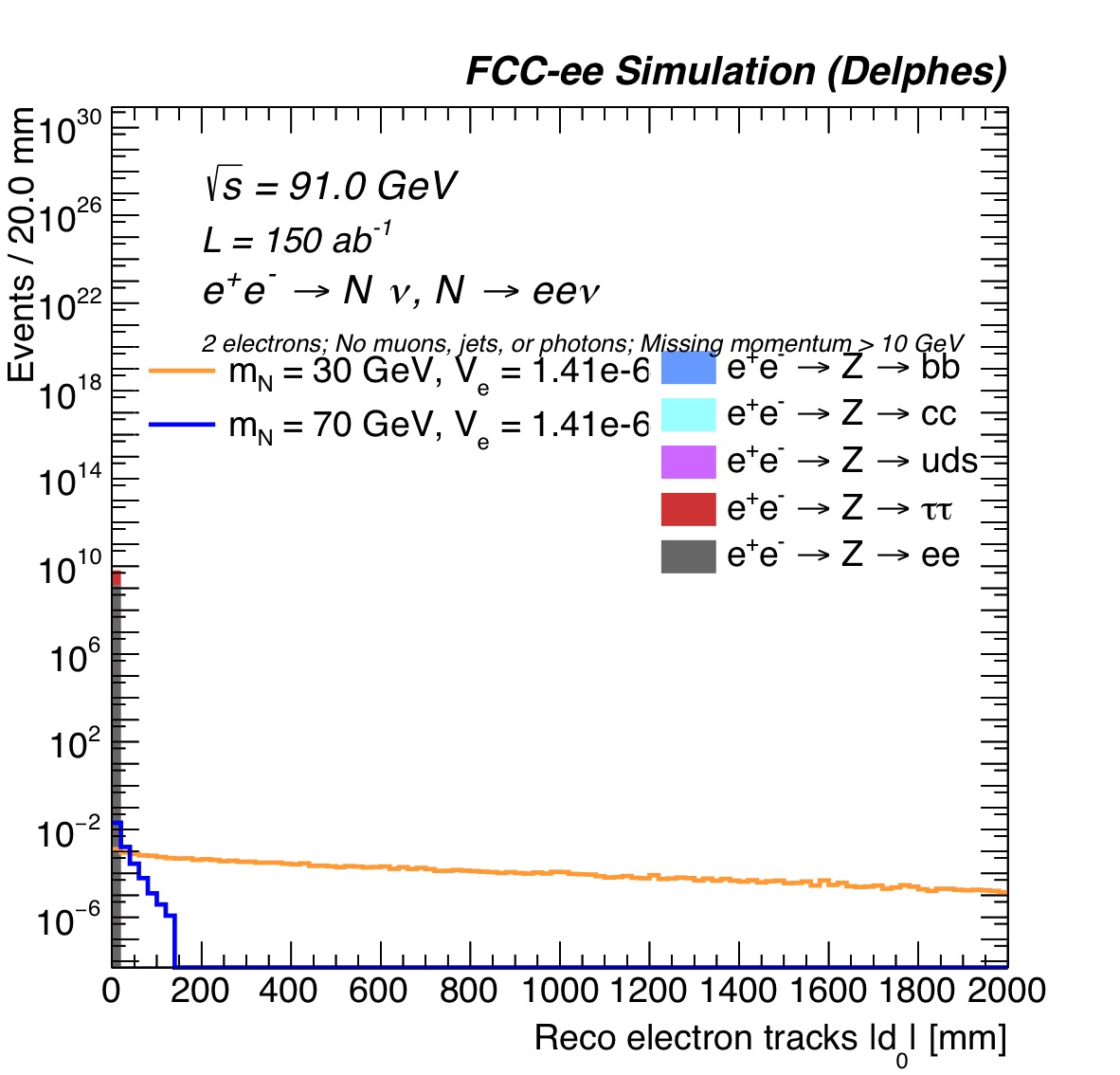}}
\caption{The normalized, reconstructed-level absolute value of the transverse impact parameter $|d_0|$, for representative HNL signal benchmark mass and $\vert V_{eN}\vert$ choices, as well as background processes, for (a) 0--1~mm in $|d_0|$ and (b) 0--2000~mm in $|d_0|$. Exactly two reconstructed electrons are required, as well as that there are no reconstructed muons, jets or photons in each event. The total missing momentum must be greater than 10~GeV.
}
\label{fig:HNLd0}
\end{figure}

Figures~\ref{fig:HNLmissingEnergy} and~\ref{fig:HNLd0} show distributions of variables that distinguish the HNL signal from these background processes. Figure~\ref{fig:HNLmissingEnergy} shows the total missing momentum $\not\! p$ in each event. Unlike in a hadron collider, where only the missing momentum in the transverse direction can be considered, the three-dimensional missing momentum can be used at FCC-ee. As can be seen from this figure, requiring $\not\! p > 10$~GeV significantly reduces the background contributions while maintaining a high efficiency for the HNL signal.

Figure~\ref{fig:HNLd0} shows the electron-track transverse impact parameter $|d_0|$ for each event. The transverse impact parameter is the distance of closest approach in the transverse plane of the helical trajectory of the track with respect to the beam axis; it is a measurement of the reconstructed electron's displacement. Requiring that both electron tracks have $|d_0|>0.5$ mm removes the vast majority of the background.

Taking these and other distributions into account, a simple event selection is developed, using reconstructed-level variables. Events must have exactly two electrons, and no photons, jets, or muons. These requirements substantially reduce the background from light and heavy quarks. We next require $\not\! p > 10$~GeV, which is particularly effective at reducing $Z\xrightarrow{}ee$ events with spurious missing momentum associated with finite detector resolution. Finally, we require that both electrons are displaced with $|d_0|>0.5$ mm to remove the vast majority of the remaining (prompt) backgrounds.

Table~\ref{tab:HNLbkgEventSelection} shows the expected number of background events for each cumulative selection criterion, and Table~\ref{tab:HNLsignalEventSelection} shows the same for representative HNL signal benchmark masses and $\vert V_{eN}\vert$ choices, assuming an integrated luminosity of 150 ab$^{-1}$. Within these limitations, these tables show that after all the selection criteria are applied, the background can be substantially reduced while the majority of the signal events are retained. After all the selection criteria are applied, we can expect about 1 event for an HNL with a mass of 50~GeV and $\vert V_{eN}\vert = 6\times 10^{-6}$, with an integrated luminosity of 150 ab$^{-1}$. This benchmark point is illustrative of the maximum sensitivity to long-lived HNLs that can be achieved at FCC-ee, with the current study.

\begin{table}[!t] 
    \centering 
    \caption{The expected number of events at an integrated luminosity of 150 ab$^{-1}$ is shown for the background processes, for each selection criterion. The cumulative number of events is shown. Only statistical uncertainty is taken into account.} 
    \resizebox{\textwidth}{!}{ 
    \begin{tabular}{|l|c|c|c|c|c|} \hline 
        & Before selection & Exactly 2 reco e & Vetoes & $\not\! p >$ 10 GeV & $|d_0| >$ 0.5 mm \\
        \hline
        Z $\rightarrow$ ee & $2.19\times 10^{11} \pm 6.94\times 10^{7}$ & $1.75\times 10^{11} \pm 6.19\times 10^{7}$ & $1.53\times 10^{11} \pm 5.80\times 10^{7}$ & $7.07\times 10^{8} \pm 3.94\times 10^{6}$ & $\leq 3.94\times 10^{6}$ \\

        Z $\rightarrow$ bb & $9.97\times 10^{11} \pm 4.14\times 10^{7}$ & $5.64\times 10^{8} \pm 9.85\times 10^{5}$ & $3.25\times 10^{5} \pm 2.36\times 10^{4}$ & $1.22\times 10^{5} \pm 1.45\times 10^{4}$ & $1.72\times 10^{3} \pm 1.72\times 10^{3}$ \\
        
        Z $\rightarrow$ $\tau \tau$ & $2.21\times 10^{11} \pm 7.00\times 10^{7}$ & $5.49\times 10^{9} \pm 1.10\times 10^{7}$ & $5.10\times 10^{9} \pm 1.06\times 10^{7}$ & $2.52\times 10^{9} \pm 7.47\times 10^{6}$ & $6.64\times 10^{4} \pm 3.84\times 10^{4}$ \\
        
        Z $\rightarrow$ cc & $7.82\times 10^{11} \pm 2.61\times 10^{7}$ & $1.69\times 10^{7} \pm 1.21\times 10^{5}$ & $5.22\times 10^{3} \pm 2.13\times 10^{3}$ & $1.74\times 10^{3} \pm 1.23\times 10^{3}$ & $\leq 1.23\times 10^{3}$ \\
        
        Z $\rightarrow$ uds & $2.79\times 10^{12} \pm 8.83\times 10^{7}$ & $2.30\times 10^{7} \pm 2.54\times 10^{5}$ & $2.79\times 10^{3} \pm 2.79\times 10^{3}$ & $\leq 2.79\times 10^{3}$ & $\leq 2.79\times 10^{3}$ \\
        \hline 
    \end{tabular}} 
    \label{tab:HNLbkgEventSelection} 
\end{table}

\begin{table}[!t]
\centering 
\caption{The expected number of events at an integrated luminosity of 150 ab$^{-1}$ is shown for representative HNL signal benchmark masses and $\vert V_{eN}\vert$ choices, for each selection criterion. The cumulative number of events is shown. Only statistical uncertainty is taken into account.} 
    \resizebox{\textwidth}{!}{ 
    \begin{tabular}{|l|c|c|c|c|c|} \hline
        & Before selection & Exactly 2 reco e & Vetoes & $\not\! p >$ 10 GeV & $|d_0| >$ 0.5 mm \\
        \hline
        $m_N =$ 10 GeV, $|V_{eN}| =  2\times 10^{-4}$ & 2534 $\pm$ 11 & 1006 $\pm$ 7 & 996 $\pm$ 7 & 951 $\pm$ 7 & 907 $\pm$ 7 \\
        
        $m_N =$ 20 GeV, $|V_{eN}| =  9\times 10^{-5}$ & 458 $\pm$ 2 & 313 $\pm$ 2 & 308 $\pm$ 2 & 293 $\pm$ 2 &  230 $\pm$ 1 \\
        
        $m_N =$ 20 GeV, $|V_{eN}| =  3\times 10^{-5}$ & 51.0 $\pm$ 0.2 & 34.7 $\pm$ 0.2 & 34.2 $\pm$ 0.2 & 32.6 $\pm$ 0.2 & 31.2 $\pm$ 0.2 \\
        
        $m_N =$ 30 GeV, $|V_{eN}| =  1\times 10^{-5}$ & 5.01 $\pm$ 0.02 & 3.85 $\pm$ 0.02 & 3.76 $\pm$ 0.02 & 3.54 $\pm$ 0.02 & 3.39 $\pm$ 0.02 \\
        
        $m_N =$ 50 GeV, $|V_{eN}| =  6\times 10^{-6}$ & 1.23 $\pm$ 0.01 & 0.99 $\pm$ 0.01 & 0.96 $\pm$ 0.01 & 0.92 $\pm$ 0.01 & 0.729 $\pm$ 0.004 \\ 
        \hline 
    \end{tabular}} 
    \label{tab:HNLsignalEventSelection} 
\end{table}

\FloatBarrier

\subsubsection{Majorana and Dirac Nature of the HNL}~\label{sec:md}

If HNLs exist in nature, a chief goal is to ascertain whether they are Dirac or Majorana fermions. As discussed in Section~\ref{sec:theory_nu} and elsewhere~\cite{Schechter:1981bd,Hirsch:2006yk,Duerr:2011zd,Moffat:2017feq}, determining this is tantamount to observing processes that are mediated by $N$ and exhibit LNV. However, at FCC-ee, the net lepton numbers of the processes $e^+e^- \to N\nu_\ell + N\overline{\nu_\ell}$ with $N\to ({\rm anything})$ are hidden because light neutrinos are not detected. This implies other metrics, such as angular distributions, are needed to disentangle the situation when lepton number violating states cannot be unambiguously identified. 

\begin{figure}[!t]
\centering
\subfigure[]{\includegraphics[width=0.45\textwidth]{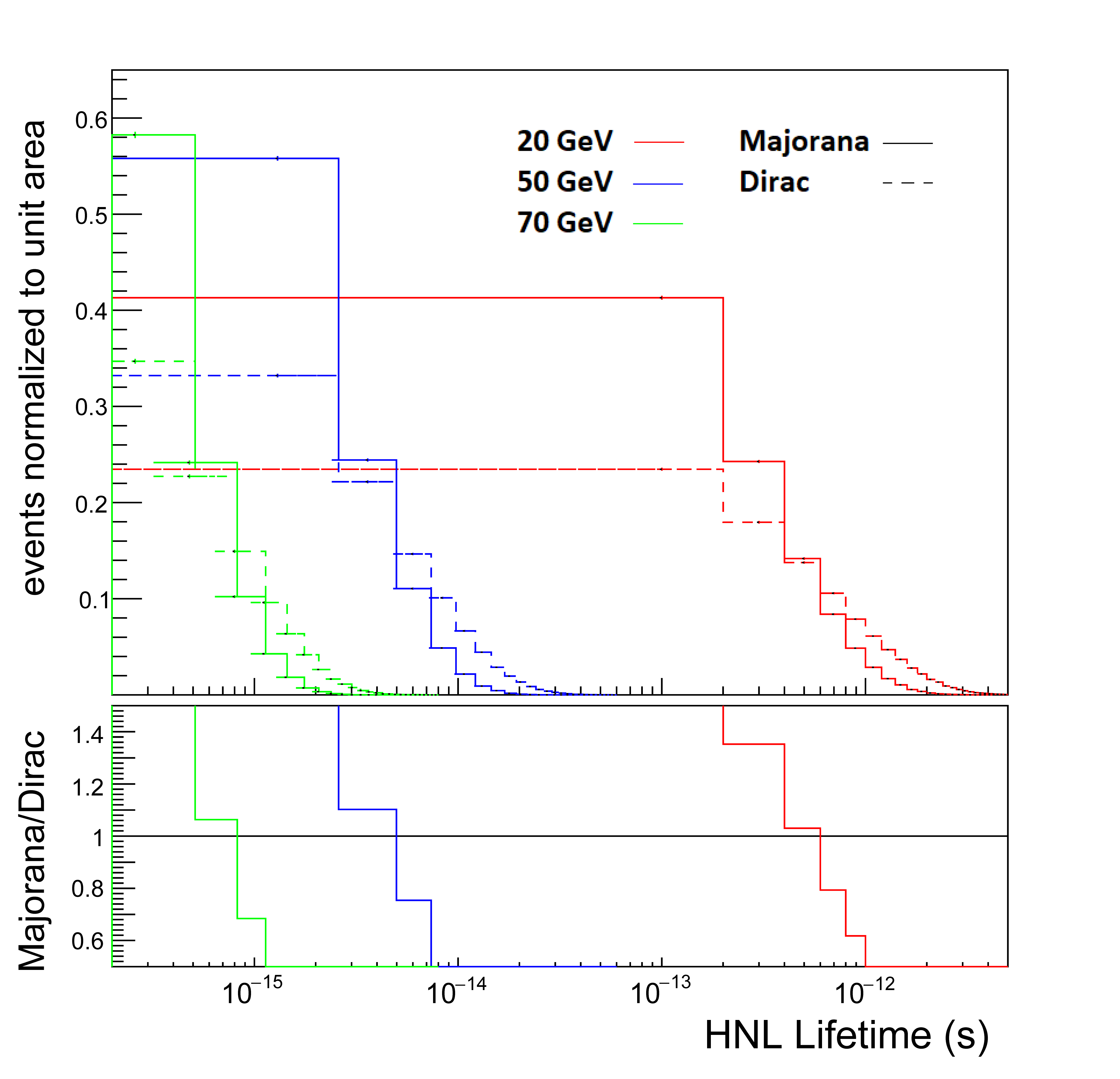} \label{fig:hnlLT}}
\subfigure[]{\includegraphics[width=0.45\textwidth]{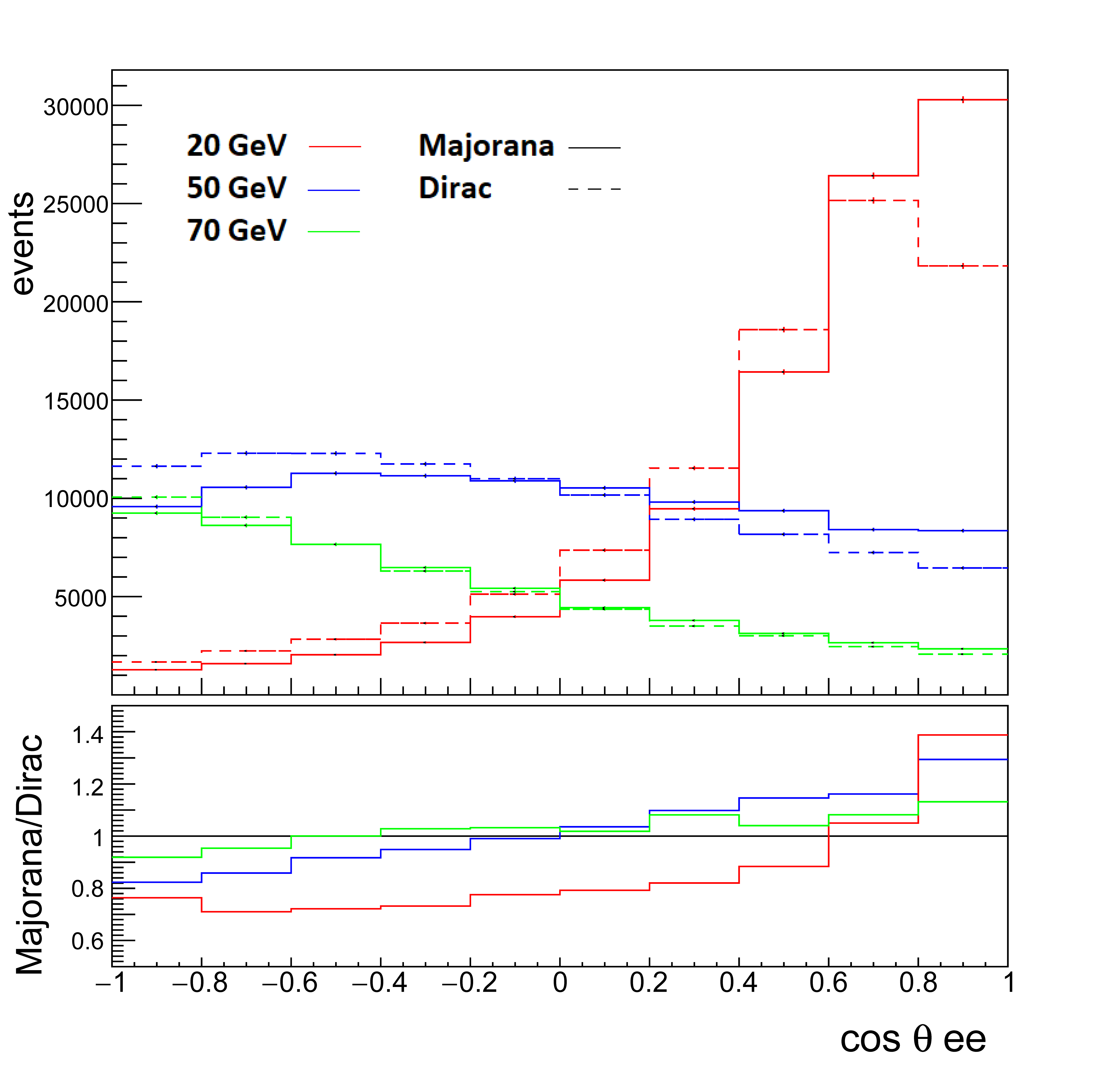} \label{fig:angsep_gen}}
\subfigure[]{\includegraphics[width=0.45\textwidth]{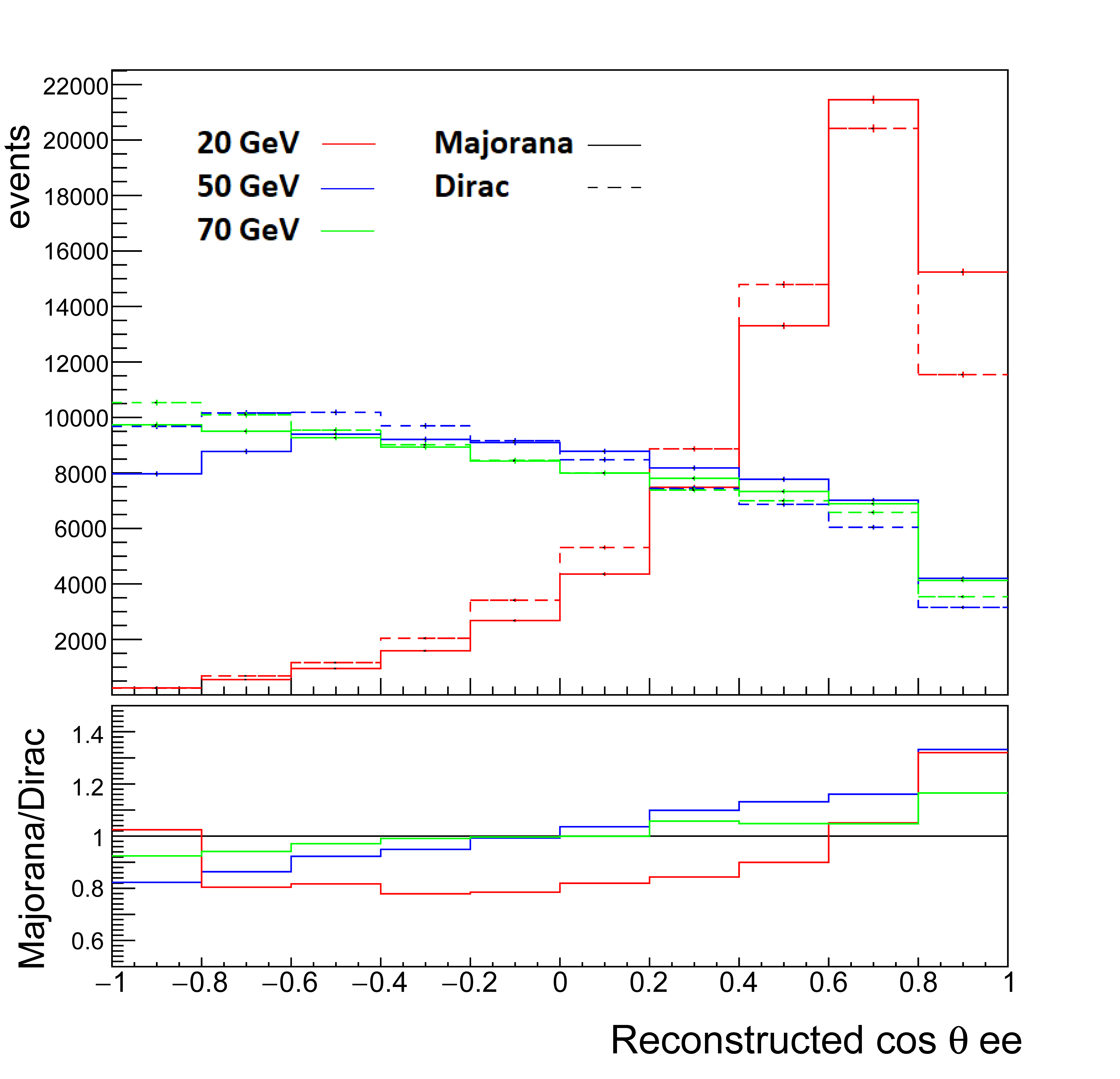}\label{fig:angsep_reco}}
\caption{
(a) The normalized distribution of lifetimes of Dirac (dashed) and Majorana (solid) HNLs in the processes defined in Eq.~\eqref{eq:sim_HNL}, for representative masses and assuming $\vert V_{e N}\vert = 10^{-3}$.
(b) The generator-level angular separation $\cos{\theta_{ee}}$ for Dirac and Majorana HNLs under the same scenario.
(c) Same as (b) but at the reconstruction level.
}
\label{fig:angsep}
\end{figure}

To demonstrate the ability of FCC-ee to potentially disentangle the Dirac or Majorana nature of HNLs, the cleanest fully-leptonic decay channels are studied; the semileptonic decay channels have about twice as large a branching ratio and will be considered in future studies. Figure~\ref{fig:angsep} shows the comparison of generator- and reconstruction-level observables for the two processes defined in Eq.~\eqref{eq:sim_HNL}. An important distinction to reiterate is that the Majorana HNL channel (solid line) includes final states that are both lepton number-conserving $(e^+e^-\nu_e\overline{\nu_e})$ as well as final states that are lepton number-violating $(e^+e^-\nu_e\nu_e, ~ e^+e^-\overline{\nu_e}\overline{\nu_e})$. On the other hand, the Dirac HNL channel  (dashed line) consists only of final states that are lepton number-conserving $(e^+e^-\nu_e\overline{\nu_e})$. Therefore, kinematical differences amount to differences between LNV and LNC. 

Figure~\ref{fig:hnlLT} shows the normalized distribution of lifetimes of Dirac and Majorana HNLs for representative masses and assuming  $\vert V_{e N}\vert = 10^{-3}$. Systematically, the lifetimes for Dirac $N$ are twice as large as for the Majorana case. For $m_N = 20-70$~GeV, the lifetimes are roughly $\tau \sim \mathcal{O}(10^{-11})-\mathcal{O}(10^{-15})$~s. As shown in equations~\eqref{eq:ZBRvN} and~\eqref{eq:Ndecayrate}, the lifetime measurement can be used together with the total cross section to distinguish between the Dirac or Majorana nature of the observed HNL, because the combination of mixing angles that appears in both quantities is the same. For this to be done correctly, two more conditions must be met. First, the mass of the HNL must be known;  this can be done by direct reconstruction, possibly combined with a kinematic fit if the HNL decays within the good quality tracker and calorimeter volumes. For longer lifetimes, a combination of decay length and laboratory decay time should be sufficient. Second, the event selection must have similar and well understood efficiencies for the three lepton flavors e, $\mu$, $\tau$, so that the differences can be corrected. 

For the same scenario, Figure~\ref{fig:angsep_gen} shows the angular separation $\cos{\theta_{ee}}$ of the $e^+e^-$ pair at the generator level. Here, several features can be observed. First, for small (large) $m_N$, the $e^+e^-$ pair are largely collimated (back-to-back). This behavior can be understood from the kinematics: a heavier $N$ is produced with less three-momentum, leading to three-body decays that are more isotropically distributed, whereas a lighter $N$ is produced with more energy, which leads to more collimated decay products.
The second feature that can be observed is that differences between the Majorana channel (LNC+LNV) and the Dirac channel (LNC) can reach $\mathcal{O}(\pm30\%)$. Differences are largest when the $e^+e^-$ pair are collimated $(\cos{\theta_{ee}}\approx 1)$ or back-to-back $(\cos{\theta_{ee}}\approx -1)$, and are smallest when they are orthogonal $(\cos{\theta_{ee}}\approx 0)$.

Finally, Figure~\ref{fig:angsep_reco} shows the same angular separation at the reconstruction level. 
Again, several features can be observed. First is that reconstruction requirements markedly impact the $\cos{\theta_{ee}}$. In particular, isolation requirements significantly reduce cases where the $e^+e^-$ pair are collimated $(\cos{\theta_{ee}}\approx 1)$. Overall, the distribution for $m_N=50$~GeV and $m_N=70$~GeV become essentially indistinguishable. Moreover, differences between the Majorana channel (LNC+LNV) and the Dirac channel (LNC) can reduce to the $\mathcal{O}(\pm20\%)$ level.

\FloatBarrier

\subsection{Axion-Like Particles}\label{sec:analysis_alp}

\begin{figure}[hbtp]
\centering
\subfigure[]{\includegraphics[width=0.45\textwidth]{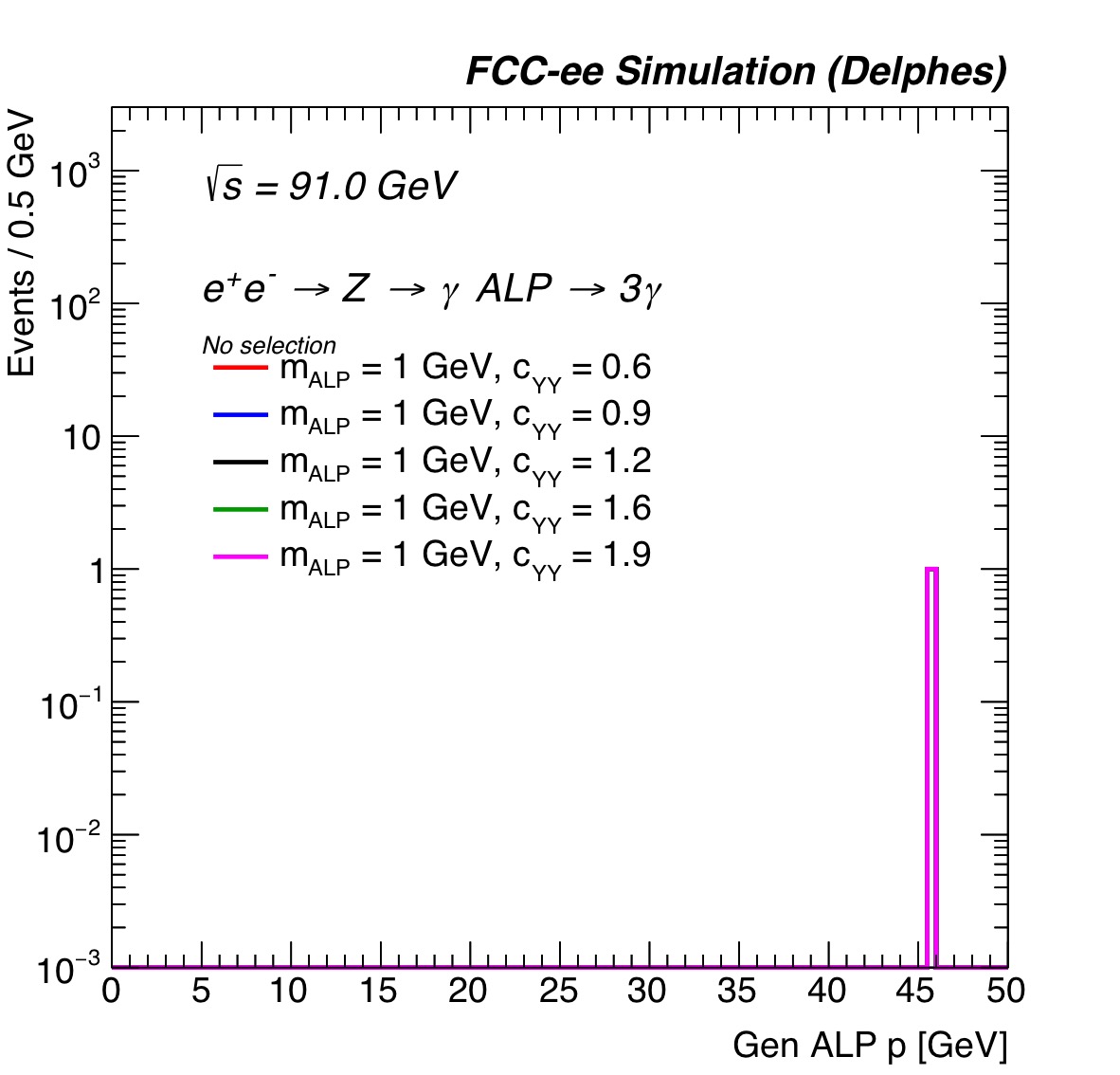}}
\subfigure[]{\includegraphics[width=0.45\textwidth]{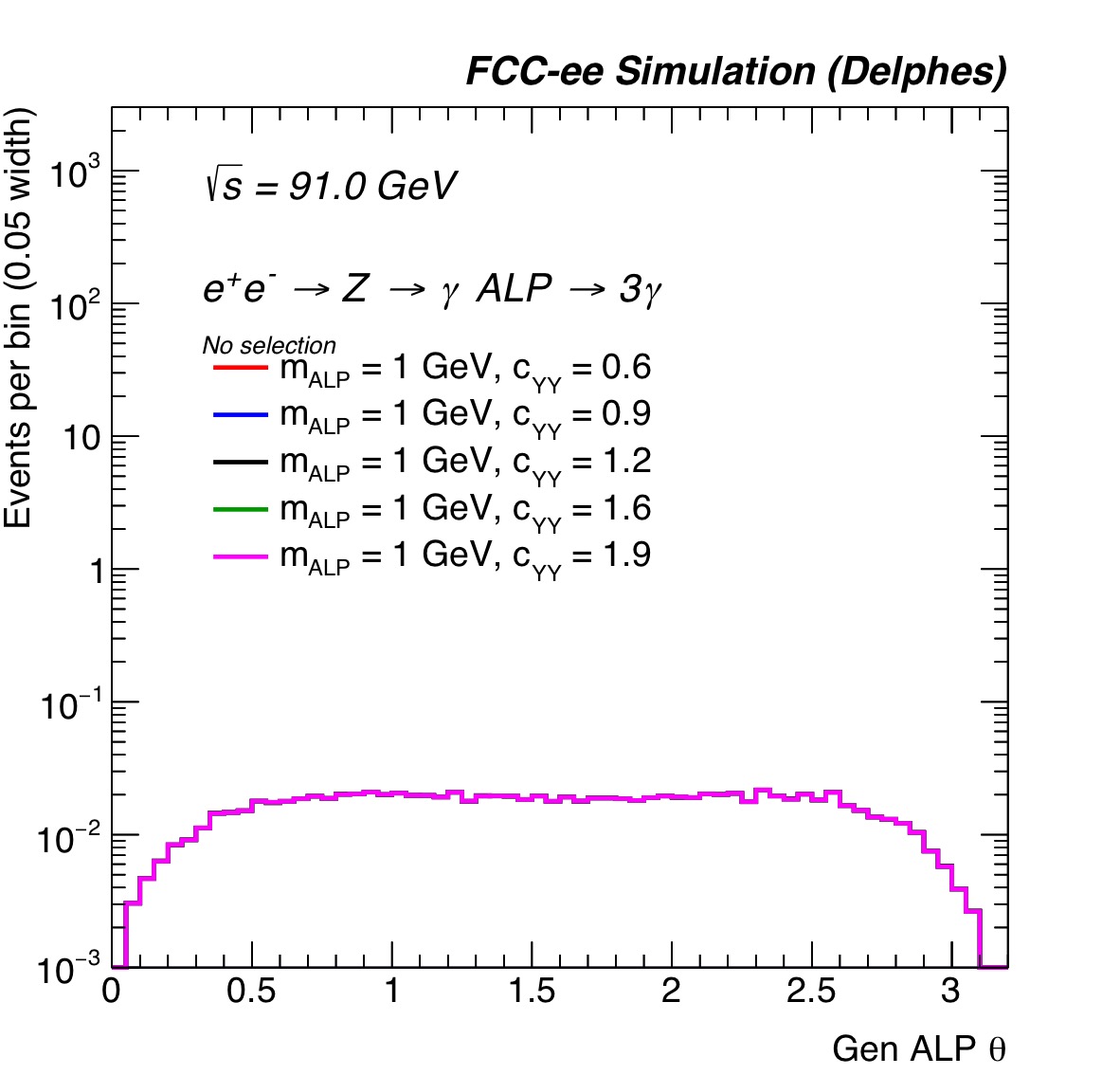}}
\caption{(a) Generated ALP momentum and (b) $\theta$ for $m_{\text{ALP}}= 1$~GeV and several benchmark choices of $c_{YY}$. The distributions are normalized to unit area.}
\label{fig:ALPkinematics}
\end{figure}

\begin{figure}[hbtp]
\centering
\subfigure[]{\includegraphics[width=0.45\textwidth]{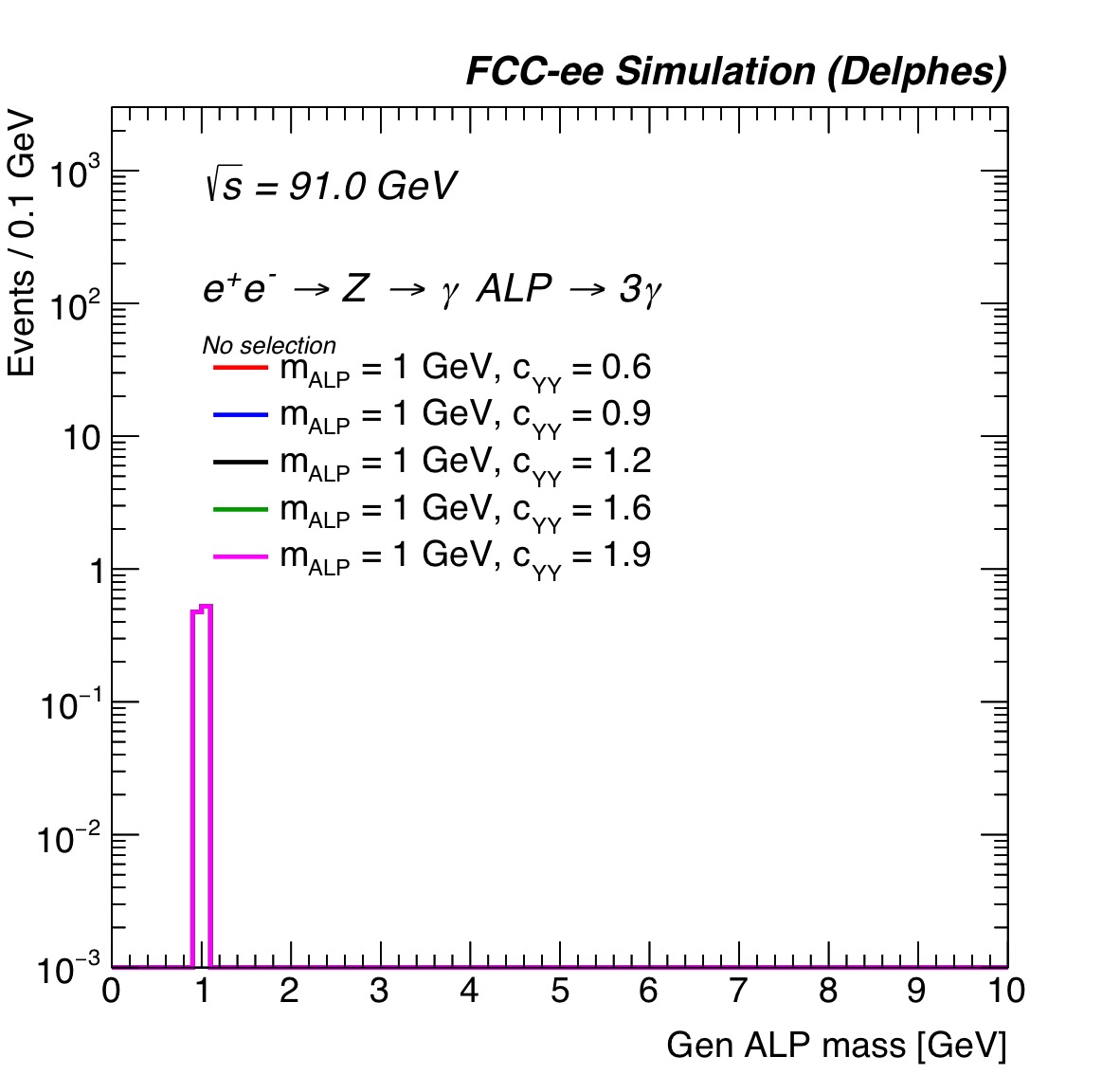}}
\subfigure[]{\includegraphics[width=0.45\textwidth]{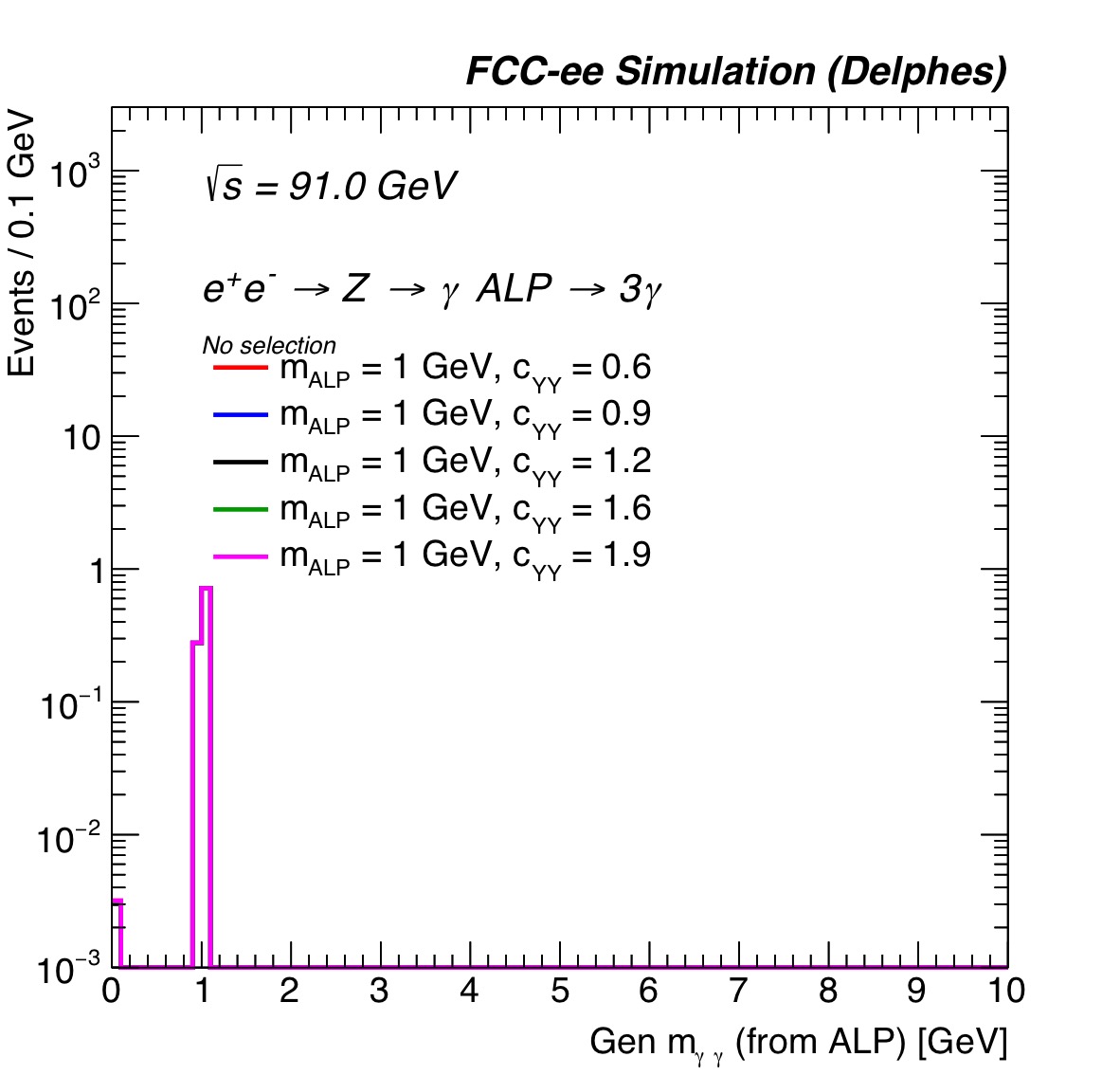}}
\subfigure[]{\includegraphics[width=0.45\textwidth]{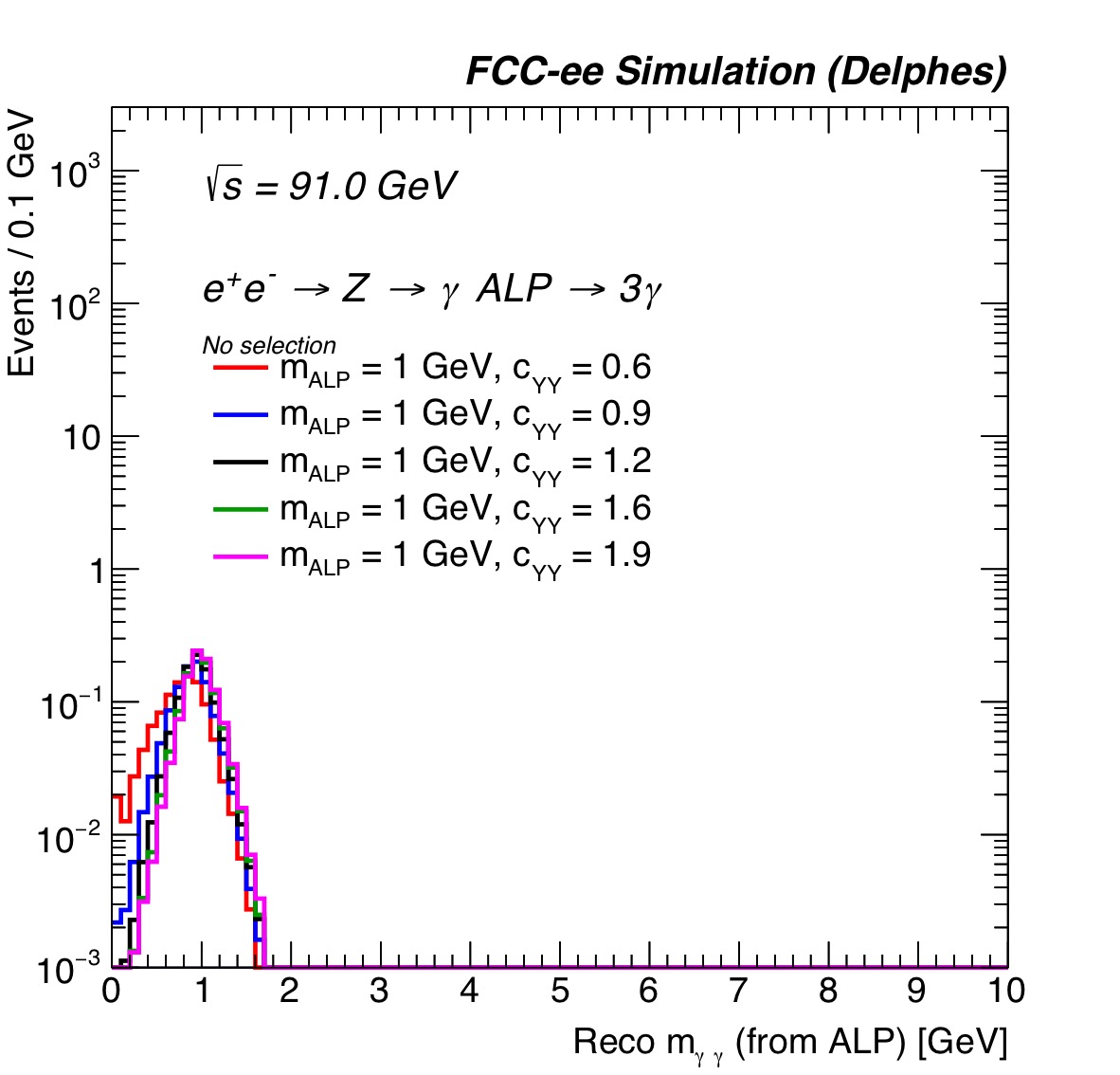}}
\caption{(a) Generated $m_{\text{ALP}}$, (b) generated $m_{\gamma\gamma}$, and (c) reconstructed $m_{\gamma\gamma}$ for $m_{\text{ALP}}= 1$~GeV and several benchmark choices of $c_{YY}$. The distributions are normalized to unit area.}
\label{fig:ALPmass}
\end{figure}

\begin{figure}[hbtp]
\centering
\subfigure[]{\includegraphics[width=0.45\textwidth]{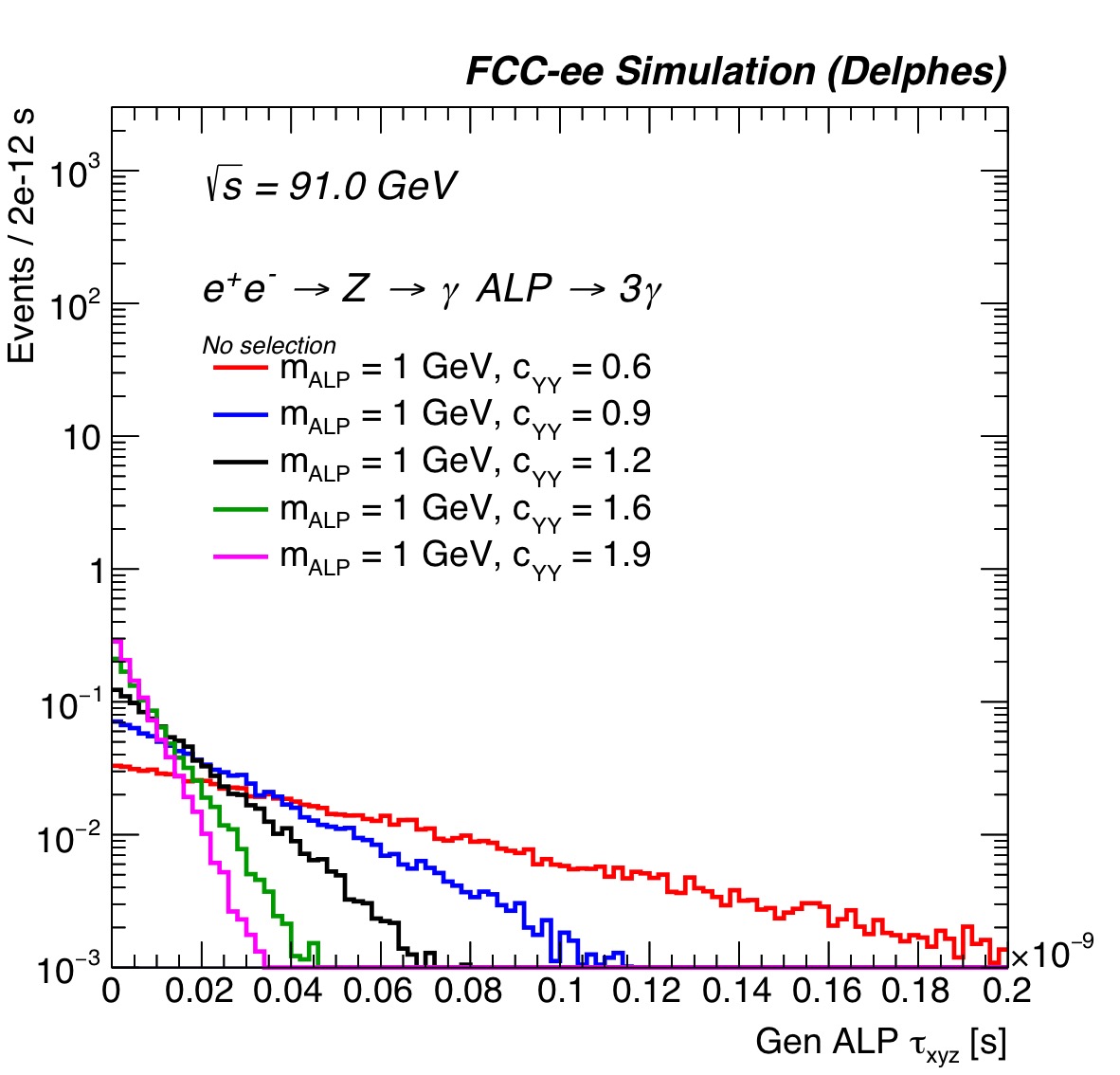}}
\subfigure[]{\includegraphics[width=0.45\textwidth]{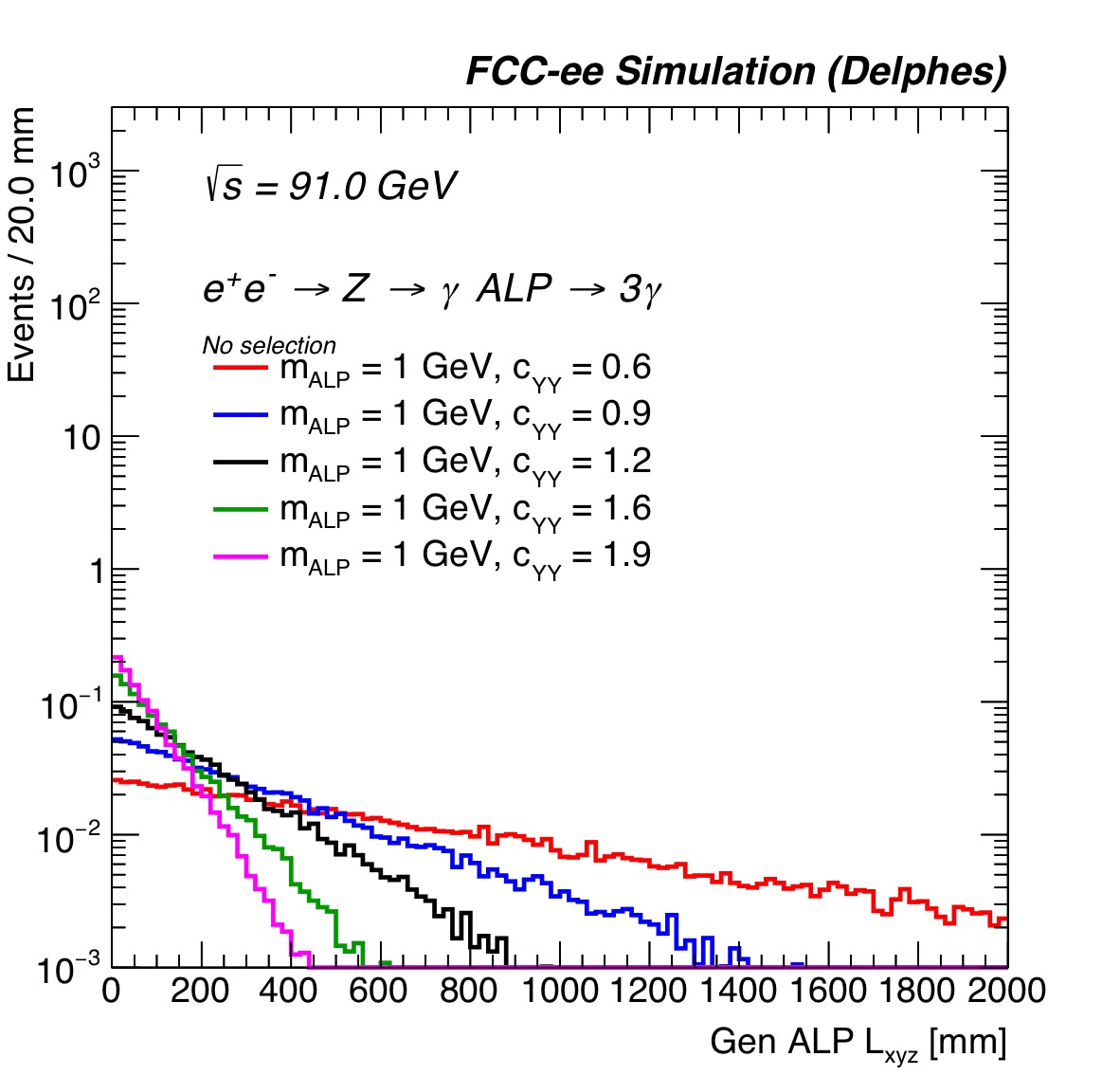}}
\caption{(a) Generated ALP $\tau_{xyz}$ and (b) $L_{xyz}$ for $m_{\text{ALP}}= 1$~GeV and several benchmark choices of $c_{YY}$. The distributions are normalized to unit area.}
\label{fig:ALPlifetime}
\end{figure}

Figure~\ref{fig:ALPkinematics} shows the generated ALP kinematics for $m_{\text{ALP}}= 1$~GeV and several benchmark choices of the hypercharge coupling $c_{YY}$. Figure~\ref{fig:ALPmass} shows the generated ALP mass ($m_{\text{ALP}}$) and the invariant mass of the two-photon system ($m_{\gamma\gamma}$), and Figure~\ref{fig:ALPlifetime} shows the generated three-dimensional lifetime $\tau_{xyz}$ and decay length $L_{xyz}$ for the ALP signal. These variables are useful in distinguishing the ALP signal from background, and also for different values of the ALP mass and couplings. In addition, calorimeter and precision timing variables will be extremely helpful to include in this study of ALPs that decay to photons. We leave these studies to a later date.

\FloatBarrier 

\subsection{Exotic Higgs Boson Decays} \label{sec:analysis_higgs}

Exotic decays of Higgs bosons to LLPs are also an interesting experimental case study at FCC-ee. As was pointed out in Section~\ref{sec:theory_higgs}, hadronic final states play a significant role, and so we plan to simulate this physics benchmark in a future paper.

\subsection{Additional Detectors for Long-Lived Particles}\label{sec:analysis_additionalDetectors}

It is possible to envisage up to four FCC-ee detectors, two of which sitting in the very large caverns foreseen from the start for the subsequent hadron collider detectors. The caverns are foreseen to be deep (200--300~m) underground, considerably reducing the cosmic ray backgrounds. A detector fully optimized for this important discovery possibility can thus be considered~\cite{Chrzaszcz:2020emg,Wang:2019xvx,Tian:2022rsi}.